\RequirePackage{fix-cm}

\documentclass[smallextended]{svjour3}       % onecolumn (second format)
\usepackage{amssymb}
\usepackage{amsmath}
\smartqed  % flush right qed marks, e.g. at end of proof
\usepackage{booktabs,caption}
\usepackage[flushleft]{threeparttable}
\usepackage{graphicx}
\usepackage{pdflscape}
\usepackage[paperheight=10in,paperwidth=8in]{geometry}
\usepackage{dblfloatfix}
 \journalname{Adv. Math. Commun.}
\begin{document}

\title{On $\mathbb{Z}_{p}\mathbb{Z}_{p}[u,v]$-additive cyclic and constacyclic codes}

\titlerunning{On $\mathbb{Z}_{p}\mathbb{Z}_{p}[u,v]$-additive cyclic and constacyclic codes}        % if too long for running head

\author{Om Prakash  \and Habibul Islam  %etc.
}


\institute{Om Prakash \at
              Department of Mathematics\\
              Indian Institute of Technology Patna, Patna 801 106, India \\
              \email{om@iitp.ac.in }
              \and Habibul Islam\at
              School of Computer Science\\
              University of St Gallen, Switzerland \\
              \email{
             habibul.islam@unisg.ch }
              }

\date{Received: date / Accepted: date}
\maketitle

\begin{abstract}
For a prime $p$, let $\mathbb{Z}_{p}$ be the ring of integers modulo $p$. The $\mathbb{Z}_{p}\mathbb{Z}_{p}[u,v]$-additive cyclic codes of length $(\alpha,\beta)$ are described as $\mathbb{Z}_{p}[u,v][x]$-submodules of $\mathbb{Z}_{p}[x]/\langle x^{\alpha}-1\rangle \times \mathbb{Z}_{p}[u,v][x]/\langle x^{\beta}-1\rangle$ where $\mathbb{Z}_{p}[u,v]=\mathbb{Z}_{p}+u\mathbb{Z}_{p}+v\mathbb{Z}_{p}$ with $u^{2}=v^{2}=uv=vu=0$. In this article, we obtain the complete sets of generator polynomials, minimal generating sets for cyclic codes of length $\beta$ over $\mathbb{Z}_{p}[u,v]$ and $\mathbb{Z}_{p}\mathbb{Z}_{p}[u,v]$-additive cyclic codes of length $(\alpha,\beta),$ respectively. It is shown that the Gray image of a $\mathbb{Z}_{p}\mathbb{Z}_{p}[u,v]$-additive cyclic code of length $(\alpha,\beta)$ is either a quasi-cyclic (QC) code of length $4\alpha$ with index $4$ or a generalized QC code of length $(\alpha,\beta,\beta,\beta)$ over $\mathbb{Z}_{p}$. Moreover, some structural properties like generating polynomials, minimal generating sets of $\mathbb{Z}_{p}\mathbb{Z}_{p}[u,v]$-additive constacyclic code of length $(\alpha,p-1)$ are determined. By defining two Gray maps $\Phi_1,\Phi_2$, we show that the $\Phi_1$-Gray image and
$\Phi_2$-Gray image of $\mathbb{Z}_{p}\mathbb{Z}_{p}[u,v]$-additive $(1+u+v)$-constacyclic codes of length $(\alpha,\beta)$ are generalized QC codes of block length $(\alpha,2\beta)$ and $(\alpha,\beta,\beta),$ respectively. Finally, we present some optimal and new linear codes obtained from these classes of constacyclic codes.
\end{abstract}

\section{Introduction}
The linear codes over finite rings have received a renewed attention of many mathematicians after the works of Hammons et al. published in 1994
\cite{Hammons}. The cyclic codes are the most important class of linear codes due to their nice algebraic properties as well as for important
generalizations, like constacyclic, quasi-cyclic, skew cyclic, etc \cite{MS,SAS}. They are defined as linear codes invariant under the cyclic shift of coordinates. The condition of linearity has been relaxed recently and replaced by additivity.
Also, the definition has been enlarged to accommodate codes over mixed alphabets.
Note that every linear code is additive, but not conversely. In 2002, Bierbrauer \cite{Bierbrauer} introduced a generalized class of cyclic codes, namely additive codes which are defined as the subgroup of a commutative group. In 2010,
Borges et al. \cite{Borges10} studied $\mathbb{Z}_{2}\mathbb{Z}_{4}$-linear codes and investigated their generator matrices and duality whereas their rank and kernel are discussed in \cite{Fernandez10}.
It is worth mentioning that such codes have a successful application in the field of information hiding, namely, steganography \cite{Rifa11}. Also, Abualrub et al. \cite{Abualrub14} characterized $\mathbb{Z}_{2}\mathbb{Z}_{4}$-additive cyclic codes as
$\mathbb{Z}_{4}[x]$-submodules of $\mathbb{Z}_{2}[x]/\langle x^{r}-1\rangle \times \mathbb{Z}_{4}[x]/\langle x^{s}-1\rangle$ and derived the unique set of generators as well as minimal generating set for these codes when $s$ is odd. Again, in 2016, Borges et al. \cite{Borges16} revisited $\mathbb{Z}_2\mathbb{Z}_4$-additive cyclic codes to discuss their generators polynomials and dual codes.
Later, mixed alphabets codes in more general setup have been rapidly explored by researchers. For instance, one can see $\mathbb{Z}_2\mathbb{Z}_{2^s}$-additive in \cite{Aydogdu13}, $\mathbb{Z}_p\mathbb{Z}_{p^s}$-additive in \cite{Shi18,Yao20} and $\mathbb{Z}_{p^r}\mathbb{Z}_{p^s}$-additive in \cite{Aydogdu15b,Yao20a}, respectively.
In addition, in 2015, Aydogdu et al. \cite{Aydogdu15a} introduced the mixed alphabets codes over $(\mathbb{Z}_2,\mathbb{Z}_2[u])$. It is obligatory to mention that they have constructed mixed codes whose second component is a codeword over the single variable polynomial ring $\mathbb{Z}_2[u]$, where $u^2=0$ rather the ring of integers modulo $p^k$.
In the next year, they \cite{Aydogdu16} have investigated the structure of $\mathbb{Z}_{2}\mathbb{Z}_{2}[u]$-cyclic and constacyclic codes where $(1+u)$-constacyclic codes of length $(\alpha,\beta)$ are $\mathbb{Z}_{2}[u][x]$-submodules of $\mathbb{Z}_{2}[x]/\langle x^{\alpha}-1\rangle \times \mathbb{Z}_{2}[u][x]/\langle x^{\beta}-(1+u)\rangle$. Moreover, with a new Gray map, they obtained some optimal binary linear codes as the Gray images of $\mathbb{Z}_{2}\mathbb{Z}_{2}[u]$-cyclic codes. Subsequently, few more articles on $\mathbb{Z}_2\mathbb{Z}_2[u]$-additive cyclic and constacyclic codes have published \cite{Borges18,P. Li,Srinivasulu}. In continuation of the literature, authors \cite{Islam20} studied $\mathbb{Z}_4\mathbb{Z}_4[u]$-additive cyclic and constacyclic codes where $\mathbb{Z}_4[u]$ is a quadratic extension of $\mathbb{Z}_4$ and obtained some new $\mathbb{Z}_4$-codes. Recently, Diao and Gao \cite{Diao18} studied $\mathbb{Z}_p\mathbb{Z}_p[u]$-additive cyclic where $u^2=0$ and Diao et al. \cite{Diao19} studied $\mathbb{Z}_p\mathbb{Z}_p[v]$-additive cyclic codes where $v^2=v$ in the context of new quantum codes construction.  We would like to mention that all of these mixed alphabets studies are limited to the single variable polynomial ring as an alphabet set. In this article for any prime $p$, we introduced the mixed alphabets codes $(\mathbb{Z}_p,\mathbb{Z}_p[u,u])$ where $\mathbb{Z}_p[u,u]$ is the double variable polynomial rings with $u^2=v^2=uv=vu=0$. It is worth mentioning that mixed alphabets constacyclic codes are fresh after \cite{Aydogdu16,P. Li,Islam20}. These codes are a generalization of the codes over $\mathbb{Z}_{p}$ as well as over $\mathbb{Z}_{p}[u,v]$.  Moreover, the choice of $\mathbb{Z}_p[u,u]$ allows us to define suitable Gray maps which produce $p$-ary optimal and new linear codes. Hence, the three major contributions of the article are categorize as follows:
\begin{enumerate}
    \item The article determines the complete set of generator polynomials and their minimal spanning sets for cyclic codes of length $\beta$ over $\mathbb{Z}_p[u,v]$.
    \item Here, we characterize the $\mathbb{Z}_p\mathbb{Z}_p[u,v]$-additive cyclic and constacyclic codes by means of their generator polynomials and minimal spanning sets.
    \item By defining new Gray maps, we obtain several $p$-ary optimal and new codes as the Gray images of these codes in which some of are improved over the best-known codes \cite{Grassl}.
\end{enumerate}
The manuscript is organized as follows. Section \ref{sec2} discusses some basic facts and useful results. Section \ref{sec3} deals
the structure of cyclic codes over $\mathbb{Z}_{p}[u,v]$ while  Section \ref{sec4} finds generator polynomials and minimal generating sets
of $\mathbb{Z}_{p}\mathbb{Z}_{p}[u,v]$-additive cyclic codes. Section \ref{sec5} includes the Gray images of cyclic
and $\mathbb{Z}_{p}\mathbb{Z}_{p}[u,v]$-additive cyclic codes. Some properties of constacyclic codes over $\mathbb{Z}_{p}[u,v]$ are obtained in
Section \ref{sec6}. Section \ref{sec7} gives the structure of $\mathbb{Z}_{p}\mathbb{Z}_{p}[u,v]$-additive constacyclic codes while
Section \ref{sec8} gives Gray images of $(1+u+v)$-constacyclic and $\mathbb{Z}_{p}\mathbb{Z}_{p}[u,v]$-additive $(1+u+v)$-constacyclic codes.
In Section \ref{sec9}, we present several optimal and new linear codes from these constacyclic codes. Section \ref{sec10} concludes the manuscript and contains few doable open problems.

\section{Preliminary}\label{sec2}
Throughout the article, for a prime $p$, $\mathbb{Z}_{p}$ represents the ring of integers modulo $p$ and $\mathbb{Z}_{p}[u,v]$ denotes
the ring $\mathbb{Z}_{p}+u\mathbb{Z}_{p}+v\mathbb{Z}_{p}$ where $u^{2}=v^{2}=uv=vu=0$. Thus $\mathbb{Z}_{p}[u,v]$ is a finite
commutative unital ring of order $p^{3}$ and of characteristic $p$. Moreover, $\mathbb{Z}_{p}[u,v]$
is a local ring with unique maximal ideal $\langle u, v\rangle$ and quotient ring $\mathbb{Z}_{p}[u,v]/\langle u,v\rangle \cong \mathbb{Z}_{p}$.
We consider $\mathbb{Z}_{p}\mathbb{Z}_{p}[u,v]:=\{(c, d)\mid c \in \mathbb{Z}_{p}, d\in \mathbb{Z}_{p}[u,v] \}$. Then
$\mathbb{Z}_{p}\mathbb{Z}_{p}[u,v]$ is a commutative group under component-wise addition. Also, for
two positive integers $\alpha$ and $\beta$, it is easy to verify that
$\mathbb{Z}_{p}^{\alpha} \times \mathbb{Z}_{p}[u,v]^{\beta}=\{(c,d) ~|~ c=(c_{1},c_{2},\cdots,c_{\alpha})\in \mathbb{Z}_{p}^{\alpha},
d=(d_{1},d_{2},\cdots,d_{\beta})\in \mathbb{Z}_{p}[u,v]^{\beta} \}$ is an additive group. We called each subgroup
of $\mathbb{Z}_{p}^{\alpha} \times \mathbb{Z}_{p}[u,v]^{\beta}$ as a $\mathbb{Z}_{p}\mathbb{Z}_{p}[u,v]$-additive code of length $(\alpha, \beta)$.
Let $\mathcal{C}$ be a $\mathbb{Z}_{p}\mathbb{Z}_{p}[u,v]$-additive code of length $(\alpha, \beta)$. Then $\mathcal{C},$ being a subgroup of
$\mathbb{Z}_{p}^{\alpha} \times \mathbb{Z}_{p}[u,v]^{\beta}$, is isomorphic to a commutative structure of the form
$\mathbb{Z}_{p}^{k_{0}}\times \mathbb{Z}_{p}^{3k_{1}}\times \mathbb{Z}_{p}^{2k_{2}}\times \mathbb{Z}_{p}^{k_{3}}$ and contains $p^{k_{0}+3k_{1}+2k_{2}+k_{3}}$ codewords.
In this case, $\mathcal{C}$ is said to be a $\mathbb{Z}_{p}\mathbb{Z}_{p}[u,v]$-additive code of type $(\alpha, \beta; k_{0}, k_{1},k_{2},k_{3})$.
Now, we define a ring homomorphism $\rho: \mathbb{Z}_{p}[u,v]\rightarrow \mathbb{Z}_{p}$ by $\rho(a+ub+vc)=a$ for $a,b,c\in \mathbb{Z}_{p}$ and multiplication $\ast: \mathbb{Z}_{p}[u,v]\times \mathbb{Z}_{p}\mathbb{Z}_{p}[u,v]\rightarrow \mathbb{Z}_{p}\mathbb{Z}_{p}[u,v]$ by $z\ast (c, d)=(\rho(z)c, zd)$, for $z\in \mathbb{Z}_{p}[u,v]$. The extension of above multiplication over $\mathbb{Z}_{p}^{\alpha}\times \mathbb{Z}_{p}[u,v]^{\beta}$ is as follows: for any $z\in \mathbb{Z}_{p}[u,v]$, we have $z\ast (c,d)=(\rho(z) c_{1},\rho(z)c_{2},\cdots,\rho(z)c_{\alpha},zd_{1},zd_{2},\cdots,zd_{\beta})$ where $c=(c_{1},c_{2},\cdots,c_{\alpha})\in \mathbb{Z}_{p}^{\alpha}, d=(d_{1},d_{2},\cdots,d_{\beta})\in \mathbb{Z}_{p}[u,v]^{\beta}$. Under this extended multiplication, it is routine to check that $\mathbb{Z}_{p}^{\alpha}\times \mathbb{Z}_{p}[u,v]^{\beta}$ is a $\mathbb{Z}_{p}[u,v]$-module.

\begin{definition}
(Module version) Any non empty subset $\mathcal{C}$ of $\mathbb{Z}_{p}^{\alpha}\times \mathbb{Z}_{p}[u,v]^{\beta}$ is said to be a $\mathbb{Z}_{p}[u,v]$-additive code of length $(\alpha, \beta)$ if $\mathcal{C}$ is a $\mathbb{Z}_{p}[u,v]$-submodule of $\mathbb{Z}_{p}^{\alpha}\times \mathbb{Z}_{p}[u,v]^{\beta}$.
\end{definition}{}

\begin{definition}
Any non empty subset $\mathcal{C}$ of $\mathbb{Z}_{p}^{\alpha}\times \mathbb{Z}_{p}[u,v]^{\beta}$ is said to be an additive cyclic code of length $(\alpha, \beta)$ if
\begin{enumerate}
\item $\mathcal{C}$ is $\mathbb{Z}_{p}\mathbb{Z}_{p}[u,v]$-additive, and
\item for any $z=(c,d)=(c_{0},c_{1},\cdots,c_{\alpha-1},d_{0},d_{1},\cdots,d_{\beta-1})\in \mathcal{C}$, we have\\
$\tau(z)=(c_{\alpha-1},c_{0},\cdots,c_{\alpha-2},d_{\beta-1},d_{0},\cdots,d_{\beta-2})\in \mathcal{C}$.
\end{enumerate}
\end{definition}
The ring homomorphism $\rho$ defined above is also extended to $\mathbb{Z}_{p}[u,v][x]$ as
\begin{align*}
\rho: \mathbb{Z}_{p}[u,v][x] \longrightarrow \mathbb{Z}_{p}[x]
\end{align*}
by
\begin{align*}
\rho(\sum_{i=0}^{n}c_{i}x^{i})=\sum_{i=0}^{n}\rho(c_{i})x^{i}.
\end{align*}
Therefore, the multiplication $\ast : \mathbb{Z}_{p}[u,v][x]\times (\mathbb{Z}_{p}[x]/\langle x^{\alpha}-1\rangle \times \mathbb{Z}_{p}[u,v][x]/\langle x^{\beta}-1\rangle)$ defined by $\gamma(x)\ast (c(x),d(x))=(\rho(\gamma(x))c(x), \gamma(x)d(x))$
is well-defined. Let $R_{\alpha, \beta}=\mathbb{Z}_{p}[x]/\langle x^{\alpha}-1\rangle \times \mathbb{Z}_{p}[u,v][x]/\langle x^{\beta}-1\rangle$ and $\mathcal{C}$ be a $\mathbb{Z}_{p}\mathbb{Z}_{p}[u,v]$-additive cyclic code of length $(\alpha, \beta)$. Now, each codeword $(c,d)=(c_{0},c_{1},\cdots,c_{\alpha-1},d_{0},d_{1},\cdots,d_{\beta-1})\in \mathcal{C}$ can identify with a polynomial $(c(x),d(x))\in R_{\alpha,\beta}$ under the correspondence $(c,d)\mapsto (c(x),d(x))$ where $c(x)=c_{0}+c_{1}x+\dots+c_{\alpha-1}x^{\alpha-1}, d(x)=d_{0}+d_{1}x+\dots+d_{\beta-1}x^{\beta-1}$. In this way, we can say $R_{\alpha, \beta}=\mathbb{Z}_{p}[x]/\langle x^{\alpha}-1\rangle \times \mathbb{Z}_{p}[u,v][x]/\langle x^{\beta}-1\rangle$ is a $\mathbb{Z}_{p}[u,v][x]$-module.

\begin{theorem}\label{th basic}
Let $\mathcal{C}$ be a $\mathbb{Z}_{p}\mathbb{Z}_{p}[u,v]$-additive code of length $(\alpha, \beta)$. Then $\mathcal{C}$ is a $\mathbb{Z}_{p}\mathbb{Z}_{p}[u,v]$-additive cyclic code if and only if $\mathcal{C}$ is a $\mathbb{Z}_{p}[u,v][x]$-submodule of $R_{\alpha, \beta}.$
\end{theorem}
\begin{proof}
Let $\mathcal{C}$ be a $\mathbb{Z}_{p}\mathbb{Z}_{p}[u,v]$-additive cyclic code of length $(\alpha, \beta)$. Let $r(x)\in \mathbb{Z}_{p}[u,v][x]$ and $z=(c(x),d(x))\in \mathcal{C}$ where $c(x)=c_{0}+c_{1}x+\dots+c_{\alpha-1}x^{\alpha-1}, d(x)=d_{0}+d_{1}x+\dots+d_{\beta-1}x^{\beta-1}$. Now, $x\ast (c(x),d(x))=(xc_{0}+c_{1}x^{2}+\dots+c_{\alpha-1}x^{\alpha},xd_{0}+d_{1}x^{2}+\dots+d_{\beta-1}x^{\beta})= (c_{\alpha-1}+c_{0}x+\dots+c_{\alpha-2}x^{\alpha-1},d_{\beta-1}+d_{0}x+\dots+d_{\beta-2}x^{\beta-1})$, represents the cyclic shift $\tau(z)$ of $z$. Also, $\mathcal{C}$ is a $\mathbb{Z}_{p}\mathbb{Z}_{p}[u,v]$-additive cyclic code, so $x^{i}\ast (c(x),d(x))\in \mathcal{C}$ for all integers $i\geq 0$. Hence, by linearity of $\mathcal{C}$, we can conclude that $\mathcal{C}$ is a $\mathbb{Z}_{p}[u,v][x]$-submodule of $R_{\alpha, \beta}.$

Conversely, let $\mathcal{C}$ be a $\mathbb{Z}_{p}[u,v][x]$-submodule of $R_{\alpha, \beta}$ and $z=(c(x),d(x))\in \mathcal{C}$. Since $\mathcal{C}$ is a $\mathbb{Z}_{p}[u,v][x]$-submodule of $R_{\alpha, \beta}$, so $x\ast (c(x),d(x))\in \mathcal{C}$. This shows that for any $z\in \mathcal{C}$, its cyclic shift $\tau(z)=x\ast (c(x),d(x))\in \mathcal{C}.$ Hence, $\mathcal{C}$ is a $\mathbb{Z}_{p}\mathbb{Z}_{p}[u,v]$-additive cyclic code.
\end{proof}

\begin{definition}
Let $z_{1}=(c_{0},c_{1},\cdots,c_{\alpha-1},d_{0},d_{1},\cdots,d_{\beta-1}), z_{2}=(e_{0},e_{1},\cdots,e_{\alpha-1},f_{0},f_{1},\\\cdots,f_{\beta-1})$ $\in\mathbb{Z}_{p}^{\alpha}\times \mathbb{Z}_{p}[u,v]^{\beta}$. The inner product of $z_{1}$
and $z_{2}$ is defined as $z_{1}\cdot z_{2}=(u+v)\sum_{i=0}^{\alpha-1}c_{i}e_{i}+\sum_{i=0}^{\beta-1}d_{i}f_{i}\in \mathbb{Z}_{p}[u,v]$. Also, for a $\mathbb{Z}_{p}\mathbb{Z}_{p}[u,v]$-additive cyclic code $\mathcal{C}$ of length $(\alpha,\beta)$, its
dual code is defined as $\mathcal{C}^{\perp}=\{z_{1}\in \mathbb{Z}_{p}^{\alpha}\times \mathbb{Z}_{p}[u,v]^{\beta} \mid z_{1}\cdot z_{2}=0, \forall z_{2}\in \mathcal{C} \}.$
\end{definition}

\begin{proposition}
For a $\mathbb{Z}_{p}\mathbb{Z}_{p}[u,v]$-additive cyclic code $\mathcal{C}$ of length $(\alpha,\beta)$, its dual $\mathcal{C}^{\perp}$ is a $\mathbb{Z}_{p}\mathbb{Z}_{p}[u,v]$-additive cyclic code.
\end{proposition}
\begin{proof}
Let $\mathcal{C}$ be a $\mathbb{Z}_{p}\mathbb{Z}_{p}[u,v]$-additive cyclic code of length $(\alpha,\beta)$.
Let $z_{1}=(c_{0},c_{1},\cdots,c_{\alpha-1},d_{0},\\d_{1}\cdots,d_{\beta-1})\in \mathcal{C}^{\perp}$. In order to prove $\tau(z_{1})\in \mathcal{C}^{\perp},$ let $z_{2}=(e_{0},e_{1},\cdots,e_{\alpha-1},f_{0},f_{1}\cdots,f_{\beta-1})\in \mathcal{C}$. Consider $k=lcm(\alpha,\beta)$ and $z_{3}=\tau^{k-1}(z_{2})=(e_{1},\cdots,e_{\alpha-1},e_{0},f_{1}\cdots,f_{\beta-1},f_{0})\in \mathcal{C}$. Then $z_{1}\cdot z_{3}=0$, and this gives
\begin{align*}
0&=(u+v)(c_{0}e_{1}+c_{1}e_{2}+\dots+c_{\alpha-2}e_{\alpha-1}+c_{\alpha-1}e_{0})+(d_{0}f_{1}+d_{1}f_{2}+\dots+\\
&\hspace{.5cm}d_{\beta-2}f_{\beta-1}+d_{\beta-1}f_{0})\\
&=(u+v)(c_{\alpha-1}e_{0}+c_{0}e_{1}+c_{1}e_{2}+\dots+c_{\alpha-2}e_{\alpha-1})+(d_{\beta-1}f_{0}+d_{0}f_{1}+d_{1}f_{2}+ \dots\\
&\hspace{.5cm} +d_{\beta-2}f_{\beta-1})\\
&=\tau(z_{1})\cdot z_{2}.
\end{align*}
Since, $z_{2}\in \mathcal{C}$ was arbitrary, so $\tau(z_{1})\in \mathcal{C}^{\perp}.$ Hence, $\mathcal{C}^{\perp}$ is a $\mathbb{Z}_{p}\mathbb{Z}_{p}[u,v]$-additive cyclic code of length $(\alpha,\beta)$.
\end{proof}

\section{Structure of cyclic codes over $\mathbb{Z}_{p}[u,v]$}\label{sec3}
The present section characterizes the cyclic codes of length $\beta$ over the ring $\mathbb{Z}_p[u,v]$. These structure will help to determine $\mathbb{Z}_p\mathbb{Z}_p[u,v]$-additive cyclic codes in subsequent sections.
\begin{definition}
A non empty subset $\mathcal{C}$ of $\mathbb{Z}_{p}[u,v]^{\beta}$ is said to be a linear code of length $\beta$ if $\mathcal{C}$ is a $\mathbb{Z}_{p}[u,v]$-submodule of $\mathbb{Z}_{p}[u,v]^{\beta}$.
\end{definition}

\begin{definition}
A linear code $\mathcal{C}$ of length $\beta$ over $\mathbb{Z}_{p}[u,v]$ is said to be cyclic if $\sigma(c):=(c_{\beta-1},c_{0},\\\cdots,c_{\beta-2})\in \mathcal{C}$ whenever $c=(c_{0},c_{1},\cdots,c_{\beta-1})\in \mathcal{C}$. Here, $\sigma$ is
known as the cyclic shift operator.
\end{definition}

We identify each codeword $c=(c_{0},c_{1},\cdots,c_{\beta-1})\in \mathcal{C}$ by a polynomial $c(x)=c_{0}+c_{1}x+\dots+c_{\beta-1}x^{\beta-1}\in \mathbb{Z}_{p}[u,v][x]/\langle x^{\beta}-1\rangle$ under the correspondence $$c=(c_{0},c_{1},\cdots,c_{\beta-1})\mapsto c(x)=c_{0}+c_{1}x+\dots+c_{\beta-1}x^{\beta-1}.$$ Therefore, in polynomial representation of the cyclic code $\mathcal{C}$, we conclude that a linear code $\mathcal{C}$ of length $\beta$ over $\mathbb{Z}_{p}[u,v]$ is cyclic if and only if $\mathcal{C}$ is an ideal of $\mathbb{Z}_{p}[u,v][x]/\langle x^{\beta}-1\rangle$.

\begin{lemma}[\cite{McDonald}] \label{lemm div}
Let $R$ be a finite commutative local ring and $f(x),g(x)\in R[x]$ be any two polynomials with $g(x)$ regular (i.e., $g(x)$ is not a zero divisor in $R[x]$). Then there exist $s(x),t(x)\in R[x]$ such that
\begin{align*}
f(x)=g(x)s(x)+t(x),
\end{align*}
where $t(x)=0$ or $deg(g(x))>deg(t(x))$.
\end{lemma}

Note that $\mathbb{Z}_{p}[u,v]$ is a finite commutative local ring, therefore, the Lemma \ref{lemm div} can be used accordingly.
One of the main targets of this article is to determine the complete structure of $\mathbb{Z}_{p}\mathbb{Z}_{p}[u,v]$-additive cyclic codes of length $(\alpha, \beta)$ and to achieve the goal, we first investigate the structure of cyclic codes of length $\beta$ over $\mathbb{Z}_{p}[u,v]$.

Here, $\mathbb{Z}_{p}[u,v]=\mathbb{Z}_{p}+u\mathbb{Z}_{p}+v\mathbb{Z}_{p}=\{a+ub+vc\mid a,b,c\in \mathbb{Z}_{p}\}$ where $u^{2}=v^{2}=uv=vu=0$ and let $\mathbb{Z}_{p}[u]=\mathbb{Z}_{p}+u\mathbb{Z}_{p}= \{a+ub\mid a,b\in \mathbb{Z}_{p}\}$ where $u^{2}=0$. Also, let $\mathcal{C}$ be a cyclic code of length $\beta$ over $\mathbb{Z}_{p}[u,v]$.
Define the map $\phi: \mathbb{Z}_{p}[u,v] \longrightarrow \mathbb{Z}_{p}[u]$ by $\phi(a+ub+vc)=a+ub$ mod $v$. Then $\phi$ is a homomorphism and
can be extended to the quotient ring $\mathbb{Z}_{p}[u,v][x]/\langle x^{\beta}-1\rangle$ as
\begin{align*}
\phi: \mathbb{Z}_{p}[u,v][x]/\langle x^{\beta}-1\rangle\longrightarrow \mathbb{Z}_{p}[u][x]/\langle x^{\beta}-1\rangle
\end{align*}
defined by
\begin{align*}
\phi(c_{0}+c_{1}x+\dots+c_{\beta-1}x^{\beta-1})=\phi(c_{0})+\phi(c_{1})x+\dots+\phi(c_{\beta-1})x^{\beta-1}.
\end{align*}
Therefore, $\mathcal{C}$ is an ideal of $\mathbb{Z}_{p}[u,v][x]/\langle x^{\beta}-1\rangle$. Also, $ker(\phi|_\mathcal{C})=vJ$ where $J$ is an ideal of $\mathbb{Z}_{p}[x]/\langle x^{n}-1\rangle$. As $\mathbb{Z}_{p}[x]/\langle x^{\beta}-1\rangle$ is a principal ideal ring, $ker(\phi|_\mathcal{C})=\langle vb(x)\rangle$ where $b(x)\in \mathbb{Z}_{p}[x]$ with $b(x)\mid (x^{\beta}-1)$ mod $p$. Further, $\phi(C)$ is an ideal of $\mathbb{Z}_{p}[u][x]/\langle x^{\beta}-1\rangle$. If $\beta$ is not relatively prime to $p$, then by Theorem 3.3 of \cite{singh15}, we have $\phi(\mathcal{C})=\langle g(x)+up_{1}(x), ua(x)\rangle$ where $a(x)\mid g(x)\mid (x^{\beta}-1)$ and $a (x)\mid p_{1}(x)\frac{x^{\beta}-1}{g(x)}$ mod $p$. Hence, $\mathcal{C}=\langle g(x)+up_{1}(x)+vp_{2}(x), ua(x)+vp_{3}(x), vb(x)\rangle$ where $a(x)\mid g(x)\mid (x^{\beta}-1)$, $b(x)\mid g(x)\mid (x^{\beta}-1)$, $a (x)\mid p_{1}(x)\frac{x^{\beta}-1}{g(x)}$ and $b (x)\mid p_{3}(x)\frac{x^{\beta}-1}{a(x)}$. If $\beta$ is relatively prime to $p$, then by Theorem 3.4 of \cite{singh15}, we have $\phi(\mathcal{C})=\langle g(x)+ua(x)\rangle$ where $a(x)\mid g(x)\mid (x^{\beta}-1)$ mod $p$ and hence $\mathcal{C}=\langle g(x)+ua(x)+vp_{1}(x), vb(x)\rangle$ where $a(x)\mid g(x)\mid (x^{\beta}-1)$, $b(x)\mid g(x)\mid (x^{\beta}-1)$ and $b (x)\mid p_{1}(x)\frac{x^{\beta}-1}{a(x)}$.

\begin{lemma}
 Let $\mathcal{C}=\langle g(x)+up_{1}(x)+vp_{2}(x), ua(x)+vp_{3}(x), vb(x)\rangle$ be a cyclic code of length $\beta$ over $\mathbb{Z}_{p}[u,v]$. If $g(x)=a(x)=b(x)$, then $\mathcal{C}=\langle g(x)+up_{1}(x)+vp_{2}(x)\rangle$ and $(g(x)+up_{1}(x)+up_{2}(x))\mid (x^{\beta}-1)$ in $\mathbb{Z}_{p}[u,v][x]/\langle x^{\beta}-1\rangle$.
\end{lemma}

\begin{proof}
We have $\phi(\mathcal{C})=\langle g(x)+up_{1}(x), ua(x)\rangle$. Now, $u(g(x)+up_{1}(x))=ug(x)=ua(x)\in \phi(\mathcal{C})$, we have $\phi(\mathcal{C})=\langle g(x)+up_{1}(x)\rangle$. Therefore, $\mathcal{C}=\langle g(x)+up_{1}(x)+vp_{2}(x), vb(x)\rangle$. Further, $v(g(x)+up_{1}(x)+vp_{2}(x))=vg(x)=vb(x)\in \mathcal{C}.$ Therefore, $\mathcal{C}=\langle g(x)+up_{1}(x)+up_{2}(x)\rangle.$ Now, dividing $x^{\beta}-1$ by $g(x)+up_{1}(x)+up_{2}(x)$, we get two polynomials $q(x), r(x)$ in $\mathbb{Z}_{p}[u,v][x]$ such that
\begin{align*}
x^{\beta}-1=(g(x)+up_{1}(x)+up_{2}(x))q(x)+r(x),
\end{align*}
where $r(x)=0$ or $deg(r(x))<ged(g(x)+up_{1}(x)+up_{2}(x))=deg(g(x))$. This implies
\begin{align*}
r(x)=-(g(x)+up_{1}(x)+up_{2}(x))q(x)\in C,
\end{align*}
in $\mathbb{Z}_{p}[u,v][x]/\langle x^{\beta}-1\rangle$. As $g(x)$ is the minimal degree polynomial in $\mathcal{C}$, $r(x)=0.$ Hence, $(g(x)+up_{1}(x)+up_{2}(x))\mid (x^{\beta}-1)$ in $\mathbb{Z}_{p}[u,v][x]/\langle x^{\beta}-1\rangle$.
\end{proof}

Now, by summarizing all these results, we characterize cyclic codes of length $\beta$ over $\mathbb{Z}_{p}[u,v]$ as follows.

\begin{theorem}\label{gen th 1}
Let $\mathcal{C}$ be a cyclic code of length $\beta$ over $\mathbb{Z}_{p}[u,v]$.
\begin{enumerate}
\item If $\beta$ is not relatively prime to $p$, then
\begin{enumerate}
\item $\mathcal{C}=\langle g(x)+up_{1}(x)+vp_{2}(x)\rangle$ with $(g(x)+up_{1}(x)+vp_{1}(x))\mid (x^{\beta}-1)$ in $\mathbb{Z}_{p}[u,v][x]/\langle x^{\beta}-1\rangle$, or,
\item $\mathcal{C}=\langle g(x)+up_{1}(x)+vp_{2}(x), ua(x)+vp_{3}(x), vb(x)\rangle$ where $a(x)\mid g(x)\mid (x^{\beta}-1)$, $b(x)\mid g(x)\mid (x^{\beta}-1)$, $a(x)\mid p_{1}(x)\frac{x^{\beta}-1}{g(x)}$ and $b(x)\mid p_{3}(x)\frac{x^{\beta}-1}{a(x)}$ mod $p$.
\end{enumerate}
\item If $\beta$ is relatively prime to $p$, then
$\mathcal{C}=\langle g(x)+ua(x)+vp_{1}(x), vb(x)\rangle$ where $a(x)\mid g(x)\mid (x^{\beta}-1)$, $b(x)\mid g(x)\mid (x^{\beta}-1)$ and $b(x)\mid p_{1}(x)\frac{x^{\beta}-1}{a(x)}$ mod $p$.

\end{enumerate}
\end{theorem}

\subsection{\textbf{Minimal spanning sets for cyclic codes over $\mathbb{Z}_{p}[u,v]$}}
In this section, we find the minimal generating sets for the cyclic codes of length $\beta$ over $\mathbb{Z}_{p}[u,v]$. In fact, these sets are helpful to get the generator matrices and sizes of those codes. For brevity of notation, we use $g, a$ etc to represent the polynomials $g(x), a(x)$ in rest of our discussion.

\begin{theorem}\label{min gen th1}
Let $\mathcal{C}$ be a cyclic code of length $\beta$ (not relatively prime to $p$) over $\mathbb{Z}_{p}[u,v]$.

\begin{enumerate}
\item If $\mathcal{C}=\langle g(x)+up_{1}(x)+vp_{2}(x)\rangle$ with $(g(x)+up_{1}(x)+vp_{1}(x))\mid (x^{\beta}-1)$ in $\mathbb{Z}_{p}[u,v][x]/\langle x^{\beta}-1\rangle$,
then
\begin{align*}
\Gamma = \{g+up_{1}+up_{2}, x(g+up_{1}+up_{2}),\cdots, x^{\beta-\delta-1}(g+up_{1}+up_{2}) \}
\end{align*}
is a minimal generating set of the code $\mathcal{C}$ where $deg(g+up_{1}+up_{2})=\delta.$

\item If $\mathcal{C}=\langle g(x)+up_{1}(x)+vp_{2}(x), ua(x)+vp_{3}(x), vb(x)\rangle$ where $a(x)\mid g(x)\mid (x^{\beta}-1)$ and $b(x)\mid g(x)\mid (x^{\beta}-1)$, $a(x)\mid p_{1}(x)\frac{x^{\beta}-1}{g(x)}$, $b(x)\mid p_{3}(x)\frac{x^{\beta}-1}{a(x)}$ mod $p$, then
\begin{align*}
\Gamma=&\{g+up_{1}+up_{2}, x(g+up_{1}+up_{2}),\cdots, x^{\beta-\delta-1}(g+up_{1}+up_{2}), ua+vp_{3},\\
&x(ua+vp_{3}),\cdots,x^{\delta-\gamma-1}(ua+vp_{3}), vb, x(vb),\cdots, ,x^{\delta-\epsilon-1}(vb) \}
\end{align*}
is a minimal generating set of the code $\mathcal{C}$ where $deg(g+up_{1}+vp_{2})=\delta, deg(ua+vp_{3})=\gamma, deg(vb)=\epsilon.$
\end{enumerate}
\end{theorem}

\begin{proof}

\begin{enumerate}
\item Let $c$ be a codeword in $\mathcal{C}$. Then $c=c_{1}(g+up_{1}+up_{2})$ where $c_{1}\in \mathbb{Z}_{p}[u,v][x]$. Since, $(g+up_{1}+vp_{2})\mid x^{\beta}-1$, so $x^{\beta}-1=(g+up_{1}+vp_{2})h$ for some $h\in \mathbb{Z}_{p}[u,v][x]$. If $deg(c_{1})\leq (\beta-\delta-1)$, then $c_{1}\in span(\Gamma)$. Otherwise, by the division algorithm, we have
\begin{align*}
c_{1}=hq+r,
\end{align*}
where $r=0$ or $deg(r)\leq deg(h)=(\beta-\delta-1)$. Then
\begin{align*}
c=c_{1}(g+up_{1}+vp_{2})&=(hq+r)(g+up_{1}+vp_{2})\\
&=q(x^{\beta}-1)+r(g+up_{1}+vp_{2})\\
&=r(g+up_{1}+vp_{2}).
\end{align*}
Therefore, $c\in span(\Gamma)$. Hence, the set $\Gamma$ generates the code $\mathcal{C}$. By the construction of $\Gamma$, one can easily check that none of the element of $\Gamma$ is a linear combination of remaining elements, therefore, $\Gamma$ is a minimal generating set of $\mathcal{C}$.

\item In order to show $\Gamma$ spans $\mathcal{C}$, it suffices to prove that $\Gamma$ spans the set
\begin{align*}
\Gamma'=&\{g+up_{1}+up_{2}, x(g+up_{1}+up_{2}),\cdots, x^{\beta-\delta-1}(g+up_{1}+up_{2}), ua+vp_{3}, \\
&x(ua+vp_{3}),\cdots,x^{\beta-\gamma-1}(ua+vp_{3}), vb, x(vb),\cdots, ,x^{\beta-\epsilon-1}(vb) \}.
\end{align*}
First we prove $x^{\delta-\epsilon}(vb)\in span(\Gamma)$. Since $b\mid g$, so $g(x)=k(x)b(x)$ for some $k(x)=k_{0}+k_{1}x+\dots+k_{\delta-\epsilon}x^{\delta-\epsilon}$. As $b$ and $g$ are both monic polynomials, so $k_{\delta-\epsilon}$ is a unit. Now, $x^{\delta-\epsilon}(vb)=v(k_{\delta-\epsilon})^{-1}g(x)-v(k_{\delta-\epsilon})^{-1}b(x)(k_{0}+k_{1}x+\dots+k_{\delta-\epsilon-1}x^{\delta-\epsilon-1})$. Also, $v(g+up_{1}+vp_{2})=vg\in span(\Gamma)$. Therefore, $x^{\delta-\epsilon}(vb)\in span(\Gamma)$. Similarly, it is proved that $x^{\delta-\epsilon+1}(vb), x^{\delta-\epsilon+2}(vb),\cdots, x^{\beta-\epsilon}(vb)$ are in $span(\Gamma)$. Next, we show that $x^{\delta-\gamma}(ua+vp_{3})\in span(\Gamma)$. Since $(g+up_{1}+vp_{2})$ is regular, by the division algorithm, we have
\begin{align*}
x^{\delta-\gamma}(ua+vp_{3})=(g+up_{1}+vp_{2})q+r,
\end{align*}
where $r=0$ or $deg(r)<deg(g+up_{1}+up_{2})=\delta.$ Therefore, $r=A(ua+vp_{3})+B(vb)\in \mathcal{C}$ where $A, B$ are polynomials with $deg(B)= \delta-\epsilon-1$ and $deg(A)<\delta-\gamma$, since $deg(r)<\delta$.  Hence, $x^{\delta-\gamma}(ua+vp_{3})\in span(\Gamma)$. Similarly, we can prove that
the remaining terms are also in $span(\Gamma)$. It is easy to check that none of the member of $\Gamma$ can be written as the combination of other members of $\Gamma$. Hence, $\Gamma$ is a minimal generating set for the code $\mathcal{C}$.
\end{enumerate}
\end{proof}

\begin{theorem}
Let $\mathcal{C}$ be a cyclic code of length $\beta$ (relatively prime to $p$) over $\mathbb{Z}_{p}[u,v]$. Let $\mathcal{C}=\langle g(x)+ua(x)+vp_{1}(x), vb(x)\rangle$ where $a(x)\mid g(x)\mid (x^{\beta}-1)$, $b(x)\mid g(x)\mid (x^{\beta}-1)$ and $b(x)\mid p_{1}(x)\frac{x^{\beta}-1}{a(x)}$ mod $p$. Then
\begin{align*}
\Gamma=&\{g+ua+up_{1}, x(g+ua+up_{1}),\cdots, x^{\beta-\delta-1}(g+ua+up_{1}) ,vb, x(vb),\cdots,\\
&x^{\delta-\epsilon-1}(vb) \}
\end{align*}
is a minimal generating set for the code $\mathcal{C}$ where $deg(g+up_{1}+vp_{2})=\delta, deg(vb)=\epsilon.$
\end{theorem}

\begin{proof}
Same as the proof of Theorem \ref{min gen th1}.
\end{proof}

\section{Structure of $\mathbb{Z}_{p}\mathbb{Z}_{p}[u,v]$-additive cyclic codes}\label{sec4}
Let $\mathcal{C}$ be a $\mathbb{Z}_{p}\mathbb{Z}_{p}[u,v]$-additive cyclic code of length $(\alpha, \beta)$. Then $\mathcal{C}$ is a $\mathbb{Z}_{p}[u,v][x]$-submodule of $R_{\alpha,\beta}=\mathbb{Z}_{p}[x]/\langle x^{\alpha}-1\rangle \times \mathbb{Z}_{p}[x][u,v]/\langle x^{\beta}-1\rangle$. In this section, we determine the structure of $\mathcal{C}$ as a $\mathbb{Z}_{p}[u,v][x]$-submodule of $R_{\alpha,\beta}$.
To do so, first we define the projection map
\begin{align*}
\Pi : R_{\alpha,\beta} \longrightarrow \mathbb{Z}_{p}[x][u,v]/\langle x^{\beta}-1\rangle
\end{align*}
by
\begin{align*}
\Pi(c(x),d(x))=d(x).
\end{align*}
It is checked that $\Pi$ is a $\mathbb{Z}_{p}[u,v][x]$-module homomorphism. Now, consider the restriction of $\Pi$ to the submodule $\mathcal{C}$. Then $ker(\Pi|_\mathcal{C})=\{(c,0)\in \mathcal{C}\mid c\in \mathbb{Z}_{p}[x]/\langle x^{\alpha}-1\rangle \}$. Let $I=\{c\in \mathbb{Z}_{p}[x]/\langle x^{\alpha}-1\rangle \mid (c,0)\in ker(\Pi|_\mathcal{C}) \}$. Then it is a routine work to check that $I$ is an ideal of $\mathbb{Z}_{p}[x]/\langle x^{\alpha}-1\rangle.$ Since $\mathbb{Z}_{p}[x]/\langle x^{\alpha}-1\rangle$ is a principal ideal ring, so there exists $f_{1}\in \mathbb{Z}_{p}[x]$ such that $I=\langle f_{1}\rangle$ and $f_{1}\mid (x^{\alpha}-1)$. Therefore, $ker(\Pi|_\mathcal{C})=\langle (f_{1},0)\rangle$. Moreover, $\Pi(\mathcal{C})$ is an ideal of $\mathbb{Z}_{p}[x][u,v]/\langle x^{\beta}-1\rangle$. If $\beta$ is not relatively prime to $p$, then by Part $1(b)$ of Theorem \ref{gen th 1}, we have $\Pi(\mathcal{C})=\langle g+up_{1}+vp_{2}, ua+vp_{3},vb\rangle$ with $a\mid g\mid (x^{\beta}-1)$ and $b\mid g\mid (x^{\beta}-1)$. Therefore, $\mathcal{C}=\langle (f_{1},0), (f_{2}, g+up_{1}+vp_{2}), (f_{3}, ua+vp_{3}), (f_{4}, vb)\rangle,$ for $f_{2}, f_{3}, f_{4}\in \mathbb{Z}_{p}[x]$. If we use the structure
of cyclic code over $\mathbb{Z}_{p}[u,v]$ given in Part 1(a) of Theorem \ref{gen th 1}, then $\mathcal{C}=\langle (f_{1},0), (f_{2}, g+up_{1}+vp_{2})\rangle$ where $(g+up_{1}+vp_{2})\mid (x^{\beta}-1), f_{1}\mid (x^{\alpha}-1)$ and $f_{1}, f_{2}\in \mathbb{Z}_{p}[x]$. \\
If $\beta$ is relatively prime to $p$, then similar results can be found. Thus, above discussion gives the following theorem.

\begin{theorem}\label{gen th 2}
Let $\mathcal{C}$ be a $\mathbb{Z}_{p}\mathbb{Z}_{p}[u,v]$-additive cyclic code of length $(\alpha,\beta)$.

\begin{enumerate}
\item If $\beta$ is not relatively prime to $p$, then

 \begin{enumerate}
\item $\mathcal{C}=\langle (f_{1},0), (f_{2}, g+up_{1}+vp_{2}), (f_{3}, ua+vp_{3}), (f_{4}, vb)\rangle,$ where $a\mid g\mid (x^{\beta}-1)$, $ b\mid g\mid (x^{\beta}-1), f_{1}\mid (x^{\alpha}-1) $ and $f_{i}\in \mathbb{Z}_{p}[x]$ for $i=2,3,4.$ \\
OR,
\item $\mathcal{C}=\langle (f_{1},0), (f_{2}, g+up_{1}+vp_{2})\rangle$ where $(g+up_{1}+vp_{2})\mid (x^{\beta}-1), f_{1}\mid (x^{\alpha}-1)$ and $f_{1}, f_{2}\in \mathbb{Z}_{p}[x]$.
\end{enumerate}

\item  If $\beta$ is relatively prime to $p$, then $\mathcal{C}=\langle (f_{1},0), (f_{2}, g+ua+vp_{1}), (f_{3}, vb)\rangle,$ where $a\mid g\mid (x^{\beta}-1)$, $ b\mid g\mid (x^{\beta}-1), f_{1}\mid (x^{\alpha}-1)$ and $f_{i}\in \mathbb{Z}_{p}[x]$ for $i=2,3.$
\end{enumerate}
\end{theorem}

\begin{lemma}\label{use lemma1}
 Let $\mathcal{C}$ be a $\mathbb{Z}_{p}\mathbb{Z}_{p}[u,v]$-additive cyclic code of length $(\alpha,\beta)$ given by $\mathcal{C}=\langle (f_{1},0), (f_{2}, g+up_{1}+vp_{2})\rangle$ where $(g+up_{1}+vp_{2})\mid (x^{\beta}-1)$ and $k=\frac{x^{\beta}-1}{(g+up_{1}+vp_{2})}$. Then $f_{1}\mid kf_{2}$.
\end{lemma}

\begin{proof}
Note that $\Pi[k(f_{2}, g+up_{1}+vp_{2})]=\Pi(kf_{2}, 0)=0$. Since, $(kf_{2},0)=k(f_{2}, g+up_{1}+vp_{2})\in \mathcal{C}$, so $(kf_{2},0)\in ker(\Pi)=\langle (f_{1},0)\rangle $. Hence, $f_{1}\mid kf_{2}.$
\end{proof}

\begin{lemma}\label{use lemma2}
 Let $\mathcal{C}$ be a $\mathbb{Z}_{p}\mathbb{Z}_{p}[u,v]$-additive cyclic code of length $(\alpha,\beta)$ given by $\mathcal{C}=\langle (f_{1},0), (f_{2}, g+up_{1}+vp_{2}), (f_{3}, ua+vp_{3}), (f_{4}, vb)\rangle,$ where $f_{1}\mid  (x^{\alpha}-1),a\mid g\mid (x^{\beta}-1)$, $ b\mid g\mid (x^{\beta}-1)$ and $f_{i}\in \mathbb{Z}_{p}[x]$, for $i=2,3,4$. Let $h=\frac{x^{\beta}-1}{g}, m_{1}=gcd(hp_{1},hp_{2}, x^{\beta}-1), m_{2}=\frac{x^{\beta}-1}{m_{1}},k=\frac{x^{\beta}-1}{a}, l_{1}=gcd(kp_{3}, x^{\beta}-1), l_{2}=\frac{x^{\beta}-1}{l_{1}}.$  Then
\begin{enumerate}
\item $f_{1}\mid m_{2}hf_{2}$,
\item $f_{1}\mid l_{2}kf_{3}$,
\item $f_{1}\mid \frac{x^{\beta}-1}{b}f_{4}$.
\end{enumerate}
\end{lemma}

\begin{proof}

\begin{enumerate}
\item Since $m_{1}\mid hp_{1},~ m_{1}\mid hp_{2}$, so  $hp_{1}=m_{1}k_{1}$ and $hp_{2}=m_{1}k_{2}$ for some polynomials $k_{1}, k_{2}$. Now,
\begin{align*}
\Pi(m_{2}h(f_{2},g+up_{1}+vp_{2}))&=\Pi(m_{2}hf_{2},m_{2}(uhp_{1}+vhp_{2})) \\
&=\Pi(m_{2}hf_{2},m_{2}(um_{1}k_{1}+vm_{1}k_{2}))\\
&=\Pi(m_{2}hf_{2},um_{2}m_{1}k_{1}+vm_{2}m_{1}k_{2})\\
&=\Pi(m_{2}hf_{2},0)\\
&=0.
\end{align*}
Since, $m_{2}h(f_{2},g+up_{1}+vp_{2})=(m_{2}hf_{2},0)\in \mathcal{C}$, so $(m_{2}hf_{2},0)\in ker(\Pi)=\langle (f_{1},0)\rangle$. Hence, $f_{1}\mid m_{2}hf_{2}$.

\item Since $l_{1}\mid kp_{3}$, so $kp_{3}=l_{1}l_{3}$ for some polynomial $l_{3}$. Now,
\begin{align*}
\Pi( l_{2}k(f_{3}, ua+vp_{3}))&=\Pi(l_{2}kf_{3}, vl_{2}kp_{3})\\
&=\Pi(l_{2}kf_{3},vl_{2}l_{1}l_{3})\\
&=\Pi(l_{2}kf_{3}, 0)\\
&=0.
\end{align*}
Since $l_{2}k(f_{3}, ua+vp_{3})=(l_{2}kf_{3}, 0)\in \mathcal{C}$, so $(l_{2}kf_{3}, 0)\in ker(\Pi)=\langle (f_{1},0)\rangle.$ Hence, $f_{1}\mid l_{2}kf_{3}.$

\item Note that
\begin{align*}
\Pi( \frac{x^{\beta}-1}{b}(f_{4}, vb))=\Pi(\frac{x^{\beta}-1}{b}f_{4}, 0)=0.
\end{align*}
Therefore, $(\frac{x^{\beta}-1}{b}f_{4}, 0)\in ker(\Pi)=\langle (f_{1},0)\rangle$ and hence $f_{1}\mid \frac{x^{\beta}-1}{b}f_{4}$.

\end{enumerate}
\end{proof}

\begin{theorem}\label{ad gn th 1}
Let $\mathcal{C}$ be a $\mathbb{Z}_{p}\mathbb{Z}_{p}[u,v]$-additive cyclic code of length $(\alpha,\beta)$ given by
\begin{align*}
\mathcal{C}=\langle (f_{1},0), (f_{2}, g+up_{1}+vp_{2}), (f_{3}, ua+vp_{3}), (f_{4}, vb)\rangle,
\end{align*}
where $a\mid g\mid (x^{\beta}-1)$, $ b\mid g\mid (x^{\beta}-1), f_{1}\mid (x^{\alpha}-1)$. Let $deg(f_{1})=t_{1}, deg(g)=t_{2}, deg(a)=t_{3}, deg(b)=t_{6}$ and $h=\frac{x^{\beta}-1}{g}, m_{1}=gcd(hp_{1},hp_{2}, x^{\beta}-1), m_{2}=\frac{x^{\beta}-1}{m_{1}}, deg(m_{2})=t_{4}, k=\frac{x^{\beta}-1}{a}, l_{1}=gcd(kp_{3}, x^{\beta}-1), l_{2}=\frac{x^{\beta}-1}{l_{1}},  deg(l_{2})=t_{5}$ and
\begin{align*}
S_{1}&=\bigcup_{i=0}^{\alpha-t_{1}-1}\{x^{i}*(f_{1},0) \}; \\
S_{2}&=\bigcup_{i=0}^{\beta-t_{2}-1}\{x^{i}*(f_{2},g+up_{1}+vp_{2}) \}; \\
S_{3}&=\bigcup_{i=0}^{\beta-t_{4}-1}\{x^{i}*(hf_{2},uhp_{1}+vhp_{2}) \}; \\
S_{4}&=\bigcup_{i=0}^{\beta-t_{3}-1}\{x^{i}*(f_{3},ua+vp_{3}) \}; \\
S_{5}&=\bigcup_{i=0}^{\beta-t_{5}-1}\{x^{i}*(kf_{3},vkp_{3}) \}; \\
S_{6}&=\bigcup_{i=0}^{\beta-t_{6}-1}\{x^{i}*(f_{4},vb) \}.
\end{align*}
Then $S=S_{1}\cup S_{2}\cup S_{3}\cup S_{4}\cup S_{5}\cup S_{6}$ is a minimal spanning set of the cyclic code $\mathcal{C}$. Moreover, $\mid \mathcal{C}\mid=p^{\alpha+7\beta-(t_{1}+3t_{2}+t_{3}+t_{4}+t_{5}+t_{6})}$.
\end{theorem}
\begin{proof}
Let $c$ be a codeword in $\mathcal{C}$. Then
\begin{align*}
c&=c_{1}*(f_{1},0)+c_{2}*(f_{2}, g+up_{1}+vp_{2})+c_{3}*(f_{3}, ua+vp_{3})+c_{4}*(f_{4}, vb)\\
&=(\rho(c_{1}) f_{1},0)+c_{2}*(f_{2}, g+up_{1}+vp_{2})+c_{3}*(f_{3}, ua+vp_{3})+c_{4}*(f_{4}, vb),
\end{align*}
for some $c_{i}\in \mathbb{Z}_{p}[u,v][x].$\\
If $deg(\rho(c_{1}))\leq (\alpha-t_{1}-1)$, then $c_{1}*(f_{1},0)\in span(S_{1})$. Otherwise, by the division algorithm, we have two polynomials $q, r\in \mathbb{Z}_{p}[x]$ such that
\begin{align*}
\rho(c_{1})=\frac{x^{\alpha}-1}{f_{1}}q+r,
\end{align*}
where $r=0$ or $deg(r)\leq (\alpha-t_{1}-1)$. Therefore,
\begin{align*}
(\rho(c_{1}) f_{1},0)&=((\frac{x^{\alpha}-1}{f_{1}}q+r)f_{1},0)\\
&=r(f_{1},0).
\end{align*}
As $deg(r)\leq (\alpha-t_{1}-1)$, so $(\rho(c_{1}) f_{1},0)\in span(S_{1})$.\\
Now, we show $c_{2}*(f_{2}, g+up_{1}+vp_{2})\in span(S_{1}\cup S_{2}\cup S_{3})\subset span(S)$. Again, by the division algorithm, we have
\begin{align*}
c_{2}=hq_{1}+r_{1},
\end{align*}
where $r_{1}=0$ or $deg(r_{1})\leq (\beta-t_{1}-1)$. Therefore,
\begin{align*}
c_{2}*(f_{2}, g+up_{1}+vp_{2})&=(hq_{1}+r_{1})(f_{2}, g+up_{1}+vp_{2})\\
&=q_{1}(hf_{2}, uhp_{1}+vhp_{2})+r_{1}(f_{2}, g+up_{1}+vp_{2}).
\end{align*}
Clearly $r_{1}(f_{2}, g+up_{1}+vp_{2})\in span(S_{2})$. Now, we show that $q_{1}(hf_{2}, uhp_{1}+vhp_{2})\in span(S)$. Since $m_{1}\mid hp_{2}$ and $m_{1}\mid hp_{2}$, so there exist $m_{3}, m_{4}$ such that $hp_{1}=m_{1}m_{3}, hp_{2}=m_{1}m_{4}$. Hence, $hp_{1}m_{2}=hp_{2}m_{2}=0$. By applying the division algorithm, we have
\begin{align*}
q_{1}=q_{2}m_{2}+r_{2},
\end{align*}
where $r_{2}=0$ or $deg(r_{2})\leq (\beta-t_{4}-1)$. Hence,
\begin{align*}
q_{1}(hf_{2}, uhp_{1}+vhp_{2})&=(q_{2}m_{2}+r_{2})(hf_{2}, uhp_{1}+vhp_{2})\\
&=q_{2}(m_{2}hf_{2}, uhp_{1}m_{2}+vhp_{2}m_{2})+r_{2}(hf_{2}, uhp_{1}+vhp_{2}) \\
&=q_{2}(m_{2}hf_{2}, 0)+r_{2}(hf_{2}, uhp_{1}+vhp_{2}).
\end{align*}
By Lemma \ref{use lemma2}, $f_{1}\mid m_{2}hf_{2}$, then $q_{2}(m_{2}hf_{2}, 0)\in span(S_{1})$ and also $r_{2}(hf_{2}, uhp_{1}+vhp_{2})\in span(S_{3}).$ Consequently $c_{2}*(f_{2}, g+up_{1}+vp_{2})\in span(S)$.\\
To show $c_{3}*(f_{3}, ua+vp_{3})\in span(S_{1}\cup S_{4}\cup S_{5})\subset span(S),$ we use the division algorithm and get
\begin{align*}
c_{3}=kq_{3}+r_{3},
\end{align*}
where $r_{3}=0$ or $deg(r_{3})\leq (\beta-t_{3}-1)$. This implies
\begin{align*}
c_{3}*(f_{3}, ua+vp_{3})&=(kq_{3}+r_{3})(f_{3}, ua+vp_{3})\\
&=q_{3}(kf_{3}, uak+vp_{3}k)+r_{3}(f_{3}, ua+vp_{3})\\
&=q_{3}(kf_{3}, vp_{3}k)+r_{3}(f_{3}, ua+vp_{3}).\\
\end{align*}
Clearly $r_{3}(f_{3}, ua+vp_{3})\in span(S_{4})$. Now, we prove $q_{3}(kf_{3}, vp_{3}k)\in span(S)$. Since $l_{1}\mid kp_{3}$, there exists $l_{3}$ such that $kp_{3}=l_{1}l_{3}$ and hence $kp_{3}l_{2}=0$. Again, by the division algorithm, we have
\begin{align*}
q_{3}=q_{4}l_{2}+r_{4},
\end{align*}
where $r_{4}=0$ or $deg(r_{4})\leq (\beta-t_{5}-1)$. Then
\begin{align*}
q_{3}(kf_{3}, vp_{3}k)&=(q_{4}l_{2}+r_{4})(kf_{3}, vp_{3}k)\\
&=q_{4}(l_{2}kf_{3}, vp_{3}kl_{2})+r_{4}(kf_{3}, vp_{3}k)\\
&=q_{4}(l_{2}kf_{3}, 0)+r_{4}(kf_{3}, vp_{3}k).
\end{align*}
This implies $r_{4}(kf_{3}, vp_{3}k)\in span(S_{5})$. Now, by Lemma \ref{use lemma2}, $f_{1}\mid l_{2}kf_{3}$, so $q_{4}(l_{2}kf_{3}, 0)\\\in span(S_{1})$. Consequently, $c_{3}*(f_{3}, ua+vp_{3})\in span(S)$.\\
Finally, we show $c_{4}*(f_{4}, vb)\in span(S_{1}\cup S_{6})\subset span(S).$ Further, by the division algorithm, we have
\begin{align*}
c_{4}=\frac{x^{\beta}-1}{b}q_{5}+r_{5},
\end{align*}
where $r_{5}=0$ or $deg(r_{5})\leq (\beta-t_{6}-1)$. Therefore,
\begin{align*}
c_{4}*(f_{4}, vb)&=(\frac{x^{\beta}-1}{b}q_{5}+r_{5})(f_{4}, vb)\\
&=q_{5}(\frac{x^{\beta}-1}{b}f_{4},0)+ r_{5}(f_{4}, vb).
\end{align*}
By Lemma \ref{use lemma2}, $f_{1}\mid \frac{x^{\beta}-1}{b}f_{4}$, this implies $q_{5}(\frac{x^{\beta}-1}{b}f_{4},0)\in span(S_{1})$ and also
$r_{5}(f_{4}, vb)\in span(S_{6})$. Consequently, $c_{4}*(f_{4}, vb)\in span(S)$. Hence, $S$ is a generating set of the code
$\mathcal{C}$. By the construction of $S_{i}$'s, it is easy to verify that none of the elements of $S$ is a linear combination
of the remaining elements. Thus, $S$ is a minimal spanning set for the cyclic code $\mathcal{C}$.
Note that $S_{1},S_{2},S_{3},S_{4},S_{5},S_{6} $ contribute $p^{\alpha-t_{1}},p^{3\beta-3t_{2}},p^{\beta-t_{4}},p^{\beta-t_{3}},p^{\beta-t_{5}},p^{\beta-t_{6}}$ codewords respectively and hence $\mid \mathcal{C}\mid=p^{\alpha+7\beta-(t_{1}+3t_{2}+t_{3}+t_{4}+t_{5}+t_{6})}$.
\end{proof}

\begin{lemma}\label{ad gn th 2}
 Let $\mathcal{C}$ be a $\mathbb{Z}_{p}\mathbb{Z}_{p}[u,v]$-additive cyclic code of length $(\alpha,\beta)$ given by
\begin{align*}
\mathcal{C}=\langle (f_{1}, 0), (f_{2},g+up_{1}+vp_{2})\rangle,
\end{align*}
where $(g+up_{1}+vp_{2})\mid (x^{\beta}-1)$ and $f_{1}\mid (x^{\alpha}-1)$. Let $deg(f_{1})=t_{1}, deg(g(x))=t_{2}$ and
\begin{align*}
S_{1}&=\bigcup_{i=0}^{\alpha-t_{1}-1} \{x^{i}*(f_{1},0)\};\\
S_{2}&=\bigcup_{i=0}^{\beta-t_{2}-1}\{x^{i}*(f_{2},g+up_{1}+vp_{2}) \}.
\end{align*}
Then $S=S_{1}\cup S_{2}$ is a minimal generating set of the cyclic code $\mathcal{C}$.  Moreover, $\mid \mathcal{C}\mid=p^{\alpha-t_{1}+3(\beta-t_{2})}$.
\end{lemma}

\begin{proof}
Let $c$ be a codeword of $\mathcal{C}$. Then
\begin{align*}
c&=c_{1}*(f_{1},0)+c_{2}*(f_{2}, g+up_{1}+vp_{2})\\
&=(\rho(c_{1})f_{1},0)+c_{2}*(f_{2}, g+up_{1}+vp_{2}),
\end{align*} foe some $c_{1},c_{2}\in \mathbb{Z}_{p}[u,v][x].$\\
If $deg(\rho(c_{1}))\leq (\alpha-t_{1}-1)$, then $c_{1}*(f_{1},0)\in span(S_{1})\subset span(S)$. Otherwise, the division algorithm gives
\begin{align*}
\rho(c_{1})=\frac{x^{\alpha}-1}{f_{1}}q+r,
\end{align*}
where $q, r\in \mathbb{Z}_{p}[x]$ with $r=0$ or $deg(r)\leq (\alpha-t_{1}-1)$. Therefore,
\begin{align*}
c_{1}*(f_{1},0)&=(\rho(c_{1})f_{1},0)\\
&=((\frac{x^{\alpha}-1}{f_{1}}q+r)f_{1},0)\\
&=r(f_{1},0).
\end{align*}
As $deg(r)\leq (\alpha-t_{1}-1)$, so $c_{1}*(f_{1},0)\in span(S_{1}\cup S_{2})= span(S)$.\\
Now, we show $c_{2}*(f_{2}, g+up_{1}+vp_{2})\in span(S_{2})$. Since $(g+up_{1}+vp_{2})\mid (x^{\beta}-1)$, which implies $x^{\beta}-1=k(g+up_{1}+vp_{2})$. Now, by the division algorithm, we have
\begin{align*}
c_{2}=kq_{1}+r_{1},
\end{align*}
where $r_{1}=0$ or $deg(r_{1})\leq deg(k)=(\beta-t_{2}-1)$. Therefore,
\begin{align*}
c_{2}*(f_{2}, g+up_{1}+vp_{2})&=(kq_{1}+r_{1})*(f_{2}, g+up_{1}+vp_{2})\\
&=q_{1}(kf_{2},0)+r_{1}(f_{2}, g+up_{1}+vp_{2}).
\end{align*}
By Lemma \ref{use lemma1}, $f_{1}\mid kf_{2}$, so $q_{1}(kf_{2},0)\in span(S_{1})$ and also it is clear that $r_{1}(f_{2}, g+up_{1}+vp_{2})\in span(S_{2}).$ Hence, $S$ generates the code $\mathcal{C}$. From the construction of $S_{1}$ and $S_{2}$, it is easy to see that $S$ is a minimal generating set for $\mathcal{C}$. Further, $S_{1}, S_{2}$ contribute $p^{\alpha-t_{1}}, p^{3(\beta-t_{2})}$ codewords respectively, thus, $\mid \mathcal{C}\mid =p^{\alpha-t_{1}+3(\beta-t_{2})}$.
\end{proof}

\begin{theorem}\label{ad gn th 3}
Let $\mathcal{C}$ be a $\mathbb{Z}_{p}\mathbb{Z}_{p}[u,v]$-additive cyclic code of length $(\alpha, \beta)$ given by $\mathcal{C}=\langle (f_{1},0), (f_{2}, g+ua+vp_{1}), (f_{3}, vb)\rangle,$ where $a\mid g\mid (x^{\beta}-1)$, $ b\mid g\mid (x^{\beta}-1),f_{1}\mid (x^{\alpha}-1)$ and $f_{i}\in \mathbb{Z}_{p}[x]$, for $i=2,3$. Let $h=\frac{x^{\beta}-1}{g}, m_{1}=gcd(ha, hp_{1}, x^{\beta}-1), m_{2}=\frac{x^{\beta}-1}{m_{1}}, deg(f_{1})=t_{1}, deg(g)=t_{2}, deg(m_{2})=t_{3}, deg(b)=t_{4}$ and
\begin{align*}
S_{1}&=\bigcup_{i=0}^{\alpha-t_{1}-1}\{ x^{i}\ast(f_{1},0) \};\\
S_{2}&=\bigcup_{i=0}^{\beta-t_{2}-1}\{ x^{i}\ast(f_{2},g+ua+vp_{1} ) \};\\
S_{3}&=\bigcup_{i=0}^{\beta-t_{3}-1}\{ x^{i}\ast(hf_{2},uha+vhp_{1} ) \};\\
S_{4}&=\bigcup_{i=0}^{\beta-t_{4}-1}\{ x^{i}\ast(f_{3},vb ) \}.
\end{align*}
Then $S=S_{1}\cup S_{2}\cup S_{3}\cup S_{4}$ is a minimal spanning set of the code $\mathcal{C}.$  Moreover, $\mid \mathcal{C}\mid=p^{\alpha+5\beta-(t_{1}+3t_{2}+t_{3}+t_{4})}$.
\end{theorem}

\begin{proof}
Same as the proof of Theorem \ref{ad gn th 1}.
\end{proof}

\begin{corollary}\label{ad hn cor 1}
Let $\mathcal{C}$ be a $\mathbb{Z}_{p}\mathbb{Z}_{p}[u,v]$-additive cyclic code of length $(\alpha, \beta)$ given by $\mathcal{C}=\langle (f_{1},0), (f_{2}, g+a(u+v)), (f_{3}, vb)\rangle,$ where $a\mid g\mid (x^{\beta}-1)$, $ b\mid g\mid (x^{\beta}-1),f_{1}\mid (x^{\alpha}-1)$ and $f_{i}\in \mathbb{Z}_{p}[x]$, for $i=2,3$. In other words, $\mathcal{C}$ is a $\mathbb{Z}_{p}\mathbb{Z}_{p}[u,v]$-additive cyclic code given by Theorem \ref{ad gn th 3} with $a=p_{1}$. Let $ deg(f_{1})=t_{1}, deg(g)=t_{2}, deg(a)=t_{3}, deg(b)=t_{4}$ and
\begin{align*}
S_{1}&=\bigcup_{i=0}^{\alpha-t_{1}-1}\{ x^{i}\ast(f_{1},0) \};\\
S_{2}&=\bigcup_{i=0}^{\beta-t_{2}-1}\{ x^{i}\ast(f_{2},g+(u+v)a ) \};\\
S_{3}&=\bigcup_{i=0}^{t_{2}-t_{3}-1}\{ x^{i}\ast(hf_{2},(u+v)ha ) \};\\
S_{4}&=\bigcup_{i=0}^{\beta-t_{4}-1}\{ x^{i}\ast(f_{3},vb ) \}.
\end{align*}
Then $S=S_{1}\cup S_{2}\cup S_{3}\cup S_{4}$ is a minimal spanning set of the code $\mathcal{C}.$
\end{corollary}

\begin{proof}
Note that $m_{1}=gcd(ha, hp_{1}, x^{\beta}-1)=gcd(ha, x^{\beta}-1)=ha.$ Rest part of the proof is same as the proof of Theorem \ref{ad gn th 1}.
\end{proof}

\subsection{\textbf{Encoding of $\mathbb{Z}_{p}\mathbb{Z}_{p}[u,v]$-additive cyclic codes}}
Based on Theorem \ref{ad gn th 1}, Lemma \ref{ad gn th 2}, Theorem \ref{ad gn th 3} and Corollary \ref{ad hn cor 1}, we propose an encoding algorithm for $\mathbb{Z}_{p}\mathbb{Z}_{p}[u,v]$-additive cyclic codes of length $(\alpha, \beta)$ in the next Theorem.

\begin{theorem}
Let $\mathcal{C}$ be a $\mathbb{Z}_{p}\mathbb{Z}_{p}[u,v]$-additive cyclic codes of length $(\alpha, \beta)$.

\begin{enumerate}
\item If $\mathcal{C}=\langle (f_{1},0), (f_{2}, g+up_{1}+vp_{2}), (f_{3}, ua+vp_{3}), (f_{4}, vb)\rangle,$  where $a\mid g\mid (x^{\beta}-1)$, $ b\mid g\mid (x^{\beta}-1), f_{1}\mid (x^{\alpha}-1)$ as given in Theorem \ref{ad gn th 1}, then any codeword $c(x)\in \mathcal{C}$ is encoded as
 \begin{align*}
 c(x)&=s_{1}\ast  (f_{1},0)+s_{2}\ast (f_{2},g+up_{1}+vp_{2})+s_{3}\ast (hf_{2},uhp_{1}+vhp_{2})\\
 &+s_{4}\ast (f_{3},ua+vp_{3})+s_{5}\ast (kf_{3},vkp_{3})+s_{6}\ast (f_{4},vb),
 \end{align*}
 where $s_{1},s_{3},s_{4},s_{5},s_{6}\in \mathbb{Z}_{p}[x]$ and $s_{2}\in \mathbb{Z}_{p}[u,v][x]$ with $deg(s_{1})\leq(\alpha-t_{1}-1)$, $deg(s_{i})\leq (\beta-t_{i}-1)$ for $i=2,3,\dots,6.$

 \item If $\mathcal{C}=\langle (f_{1}, 0), (f_{2},g+up_{1}+vp_{2})\rangle$, where $(g+up_{1}+vp_{2})\mid (x^{\beta}-1)$ and $f_{1}\mid (x^{\alpha}-1)$ as given in Lemma \ref{ad gn th 2}, then any codeword $c(x)\in \mathcal{C}$ is encoded as
 \begin{align*}
 c(x)&=s_{1}\ast  (f_{1},0)+s_{2}\ast (f_{2},g+up_{1}+vp_{2})
 \end{align*}
where $s_{1}\in \mathbb{Z}_{p}[x]$ and $s_{2}\in \mathbb{Z}_{p}[u,v][x]$ with $deg(s_{1})\leq (\alpha-t_{1}-1)$,  $deg(s_{2})\leq (\beta-t_{2}-1)$.

\item If $\mathcal{C}=\langle (f_{1},0), (f_{2}, g+ua+vp_{1}), (f_{3}, vb)\rangle,$ where $a\mid g\mid (x^{\beta}-1)$, $ b\mid g\mid (x^{\beta}-1),f_{1}\mid (x^{\alpha}-1)$ as given in Theorem \ref{ad gn th 3}, then any codeword $c(x)\in \mathcal{C}$ is encoded as
 \begin{align*}
 c(x)&=s_{1}\ast  (f_{1},0)+s_{2}\ast (f_{2},g+ua+vp_{1})+s_{3}\ast (hf_{2},uha+vhp_{1})\\
 &+s_{4}\ast (f_{3},vb),
 \end{align*}
 where $s_{1},s_{3},s_{4}\in \mathbb{Z}_{p}[x]$ and $s_{2}\in \mathbb{Z}_{p}[u,v][x]$ with $deg(s_{1})\leq (\alpha-t_{1}-1)$, $deg(s_{i})\leq (\beta-t_{i}-1)$ for $i=2,3,4.$

 \item If $\mathcal{C}=\langle (f_{1},0), (f_{2}, g+a(u+v)), (f_{3}, vb)\rangle,$ where $a\mid g\mid (x^{\beta}-1)$, $ b\mid g\mid (x^{\beta}-1),f_{1}\mid (x^{\alpha}-1)$ and $f_{i}\in \mathbb{Z}_{p}[x]$ as given in Corollary \ref{ad hn cor 1}, then any codeword $c(x)\in \mathcal{C}$ is encoded as
 \begin{align*}
 c(x)&=s_{1}\ast  (f_{1},0)+s_{2}\ast (f_{2},g+(u+v)a)+s_{3}\ast (hf_{2},(u+v)ha)\\
 &+s_{4}\ast (f_{3},vb),
 \end{align*}
 where $s_{1},s_{3},s_{4}\in \mathbb{Z}_{p}[x]$ and $s_{2}\in \mathbb{Z}_{p}[u,v][x]$ with $deg(s_{1})\leq (\alpha-t_{1}-1)$, $deg(s_{i})\leq (\beta-t_{i}-1)$ for $i=2,4$ and $deg(s_{3})\leq (t_{2}-t_{3}-1)$.
\end{enumerate}
\end{theorem}

\begin{proof}
Follows from Theorems \ref{ad gn th 1}, Lemma \ref{ad gn th 2}, Theorem \ref{ad gn th 3} and Corollary \ref{ad hn cor 1}.
\end{proof}

\section{Gray map and $\mathbb{Z}_{p}$-images of $\mathbb{Z}_{p}\mathbb{Z}_{p}[u,v]$-additive cyclic codes}\label{sec5}
\begin{definition}
Let $\mathcal{C}$ be a linear code of length $\beta=st$ over $\mathbb{Z}_{p}$. We define the quasi-cyclic shift operator $\pi_{s}: \mathbb{Z}_{p}^{n}\longrightarrow \mathbb{Z}_{p}^{n}$ by
\begin{align}\label{map quasi-cyclic}
\pi_{s}(e_{0}\mid e_{1}\mid\dots\mid e_{s-1})=(\sigma(e_{0})\mid \sigma(e_{1})\mid \dots\mid \sigma(e_{s-1})),
\end{align}
where $e_{i}\in \mathbb{Z}_{p}^{t}$ for all $i=0,1,\dots,(s-1)$ and $\sigma$ is the cyclic shift operator. Then $\mathcal{C}$ is said to be a quasi-cyclic (QC) code of index $s$ if $\mathcal{C}$ is invariant under the map $\pi_{s}$, i.e., $\pi_{s}(\mathcal{C})=\mathcal{C}.$
\end{definition}

We define a map $\psi:\mathbb{Z}_{p}[u,v]\longrightarrow \mathbb{Z}_{p}^3$ by $\psi(a+ub+vc)=(a,a+b,a+c)$, where $a,b,c\in \mathbb{Z}_p$.
The map $\psi$ is linear and also its extension to $\mathbb{Z}_p[u,v]^\beta$ is defined by  $\psi(r_{0},r_{1},\cdots,r_{\beta-1})=(a_0,\cdots,a_{\beta-1},a_0+b_0,\cdots,a_{\beta-1}+b_{\beta-1},a_0+c_0,\cdots,a_{\beta-1}+c_{\beta-1})$ where $r_i=a_i+ub_i+vc_i$ for $i=0,\cdots,\beta-1$. The maps $\psi$ is isometric where the Gray distance is defined as $d_{G}(r,r')=w_{G}(r-r')$ and $w_{G}(r)=\psi(w_{H}(r))$, for $r,r'\in \mathbb{Z}_{p}[u,v]^{\beta}$ and $w_H$
denotes the Hamming weight over the alphabet $\mathbb{Z}_p$.

\begin{theorem}
If $\mathcal{C}$ is a cyclic code of length $\beta$ over $\mathbb{Z}_p[u,v]$, then $\psi(\mathcal{C})$ is a quasi-cyclic code of length $3\beta$ and index $3$ over $\mathbb{Z}_p$.
\end{theorem}
\begin{proof}
It is easy to verify.
\end{proof}
Further, we define the map $\Psi: \mathbb{Z}_{p}\mathbb{Z}_{p}[u,v]\longrightarrow \mathbb{Z}_{p}^4$ by $\Psi(e,r)=\Psi(e, a+ub+vc)=(e,\psi(a+ub+vc))=(e,a,a+b,a+c)$ where $r=a+ub+vc\in \mathbb{Z}_{p}[u,v]$ and $a,b,c,e\in \mathbb{Z}_{p}$. Also, the extension of the map $\Psi$ is
\begin{align*}
\Psi : \mathbb{Z}_{p}^{\alpha}\times \mathbb{Z}_{p}[u,v]^{\beta}\longrightarrow \mathbb{Z}_{p}^{\alpha+3\beta}
\end{align*}
defined by
\begin{align*}
&\Psi(e_{0},e_{1},\cdots,e_{\alpha-1},r_{0},r_{1},\cdots,r_{\beta-1})\\
=&\Psi(e_{0},e_{1},\cdots,e_{\alpha-1},a_{0}+ub_{0}+vc_{0},a_{1}+ub_{1}+vc_{1},\cdots,a_{\beta-1}+ub_{\beta-1}+vc_{\beta-1})\\
=&(e_{0},e_{1},\cdots,e_{\alpha-1},a_{0},a_{1},\cdots,a_{\beta-1},a_{0}+b_{0},a_{1}+b_{1},\cdots,a_{\beta-1}+b_{\beta-1},a_{0}+c_{0},\\
&a_{1}+c_{1},\cdots,a_{\beta-1}+c_{\beta-1}),
\end{align*}
where $r_{i}=a_{i}+ub_{i}+vc_{i}$, for $i=0,1,\dots,(\beta-1)$.

\begin{theorem}\label{lemma gray}
Let $\mathcal{C}$ be a $\mathbb{Z}_{p}\mathbb{Z}_{p}[u,v]$-additive cyclic code of length $(\alpha, \beta)$. Then $\Psi(\mathcal{C})$ is a
\begin{enumerate}
\item QC code of length $4\alpha$ and index $4$ over $\mathbb{Z}_{p}$ if $\alpha=\beta$,
\item Generalized  QC code of block length $(\alpha,\beta,\beta,\beta)$ over $\mathbb{Z}_{p}$ if $\alpha\neq \beta$. [See the Definition 1 of \cite{Esmaeili}]
\end{enumerate}
\end{theorem}
\begin{proof}
Let $(e_{0},e_{1},\cdots,e_{\alpha-1},a_{0},a_{1},\cdots,a_{\beta-1},a_{0}+b_{0},a_{1}+b_{1},\cdots,a_{\beta-1}+b_{\beta-1},a_{0}+c_{0},a_{1}+c_{1},\cdots,a_{\beta-1}+c_{\beta-1})\in \Psi(\mathcal{C}).$ Then there exists $(e_{0},e_{1},\cdots,e_{\alpha-1},r_{0},r_{1},\cdots,\\r_{\beta-1})\in \mathcal{C}$ such that
\begin{align*}
&\Psi(e_{0},e_{1},\cdots,e_{\alpha-1},r_{0},r_{1},\cdots,r_{\beta-1})\\
=&\Psi(e_{0},e_{1},\cdots,e_{\alpha-1},a_{0}+ub_{0}+vc_{0},a_{1}+ub_{1}+vc_{1},\cdots,a_{\beta-1}+ub_{\beta-1}+vc_{\beta-1})\\
=&(e_{0},e_{1},\cdots,e_{\alpha-1},a_{0},a_{1},\cdots,a_{\beta-1},a_{0}+b_{0},a_{1}+b_{1},\cdots,a_{\beta-1}+b_{\beta-1},a_{0}+c_{0},\\
&a_{1}+c_{1},\cdots,a_{\beta-1}+c_{\beta-1}),
\end{align*}
where $r_{i}=a_{i}+ub_{i}+vc_{i}$, for $i=0,1,\cdots,(\beta-1)$. Since, $\mathcal{C}$ is $\mathbb{Z}_{p}\mathbb{Z}_{p}[u,v]$-additive cyclic,\\ so $(e_{\alpha-1},e_{0},\dots,e_{\alpha-2},r_{\beta-1},r_{0},\cdots,r_{\beta-2})\in \mathcal{C}$. Therefore,
\begin{align*}
&\Psi(e_{\alpha-1},e_{0},\cdots,e_{\alpha-2},r_{\beta-1},r_{0},\cdots,r_{\beta-2})\\
&=(e_{\alpha-1},e_{0},\cdots,e_{\alpha-2},a_{\beta-1},a_{0},\cdots,a_{\beta-2},a_{\beta-1}+b_{\beta-1},a_{0}+b_{0},\cdots,a_{\beta-2}+b_{\beta-2},\\
&\hspace{.5cm} a_{\beta-1}+c_{\beta-1},a_{0}+c_{0},\cdots,a_{\beta-2}+c_{\beta-2})\in \Psi(\mathcal{C}).
\end{align*}
This shows that
\begin{enumerate}
\item if $\alpha=\beta$, then $\Psi(\mathcal{C})$ is a QC code of length $4\alpha$ and index $4$ over $\mathbb{Z}_{p}$.
\item if $\alpha\neq \beta$, then $\Psi(\mathcal{C})$ is a generalized  QC code of block length $(\alpha, \beta,\beta,\beta)$  over $\mathbb{Z}_{p}$.
\end{enumerate}
\end{proof}

\section{Structure of constacyclic codes over $\mathbb{Z}_{p}[u,v]$}\label{sec6}
In this section, we investigate the algebraic properties of constacyclic codes over $\mathbb{Z}_{p}[u,v]$. These properties will help to find the generator polynomials of $\mathbb{Z}_{p}\mathbb{Z}_{p}[u,v]$-additive constacyclic codes.

\begin{definition}
Let $\lambda$ be a unit in $\mathbb{Z}_{p}[u,v]$. A linear code $\mathcal{C}$ of length $\beta$ is said to be a  $\lambda$-constacyclic code over $\mathbb{Z}_{p}[u,v]$ if $\Upsilon_{\lambda}(c)=(\lambda c_{\beta-1}, c_{0},\cdots, c_{\beta-2})\in \mathcal{C}$ whenever $c=( c_{0}, c_{1},\cdots, \\c_{\beta-1})\in \mathcal{C}$. Here, $\Upsilon_{\lambda}$ is known as $\lambda$-constacyclic shift operator. Note that $\mathcal{C}$ is a cyclic code for $\lambda=1$ and negacyclic code for $\lambda=-1$.
\end{definition}

We identify each codeword $c=(c_{0},c_{1},\cdots,c_{\beta-1})\in \mathcal{C}$ with a polynomial $c(x)$ in \\$\mathbb{Z}_{p}[u,v][x]/\langle x^{\beta}-\lambda \rangle$ as follows:
\begin{align*}
c=(c_{0},c_{1},\cdots,c_{\beta-1}) \mapsto c_{0}+c_{1}x+\dots+c_{\beta-1}x^{\beta-1}=c(x).
\end{align*}
By the above identification, we conclude that a linear code $\mathcal{C}$ of length $\beta$ over $\mathbb{Z}_{p}[u,v]$ is a $\lambda$-constacyclic code if and only if $\mathcal{C}$ is an ideal of $\mathbb{Z}_{p}[u,v][x]/\langle x^{\beta}-\lambda \rangle$. We choose $\lambda \in \mathbb{Z}_{p}[u,v]$ is a unit so that $\lambda^{p} =1$. Let $\lambda=\lambda_{1}+u\lambda_{2}+v\lambda_{3}$, where $\lambda_{1},\lambda_{2},\lambda_{3}\in \mathbb{Z}_{p}$. Since $\mathbb{Z}_{p}[u,v]$ is a commutative ring with characteristic $p$, we have $\lambda_{1}^{p}=1$. Also, for any non zero $\lambda_{1}\in \mathbb{Z}_{p}$, we know $\lambda_{1}^{p}=\lambda_{1}$ and hence $\lambda_{1}=1$. Therefore, $\lambda=1+u\lambda_{2}+v\lambda_{3}$. For the further calculation, the value of $\lambda=1+u\lambda_{2}+v\lambda_{3}$. Now,
for the prime $p>2$, we define a map
\begin{align*}
T: \mathbb{Z}_{p}[u,v][x]/\langle x^{p-1}-1\rangle \longrightarrow \mathbb{Z}_{p}[u,v][x]/\langle x^{p-1}-\lambda \rangle
\end{align*}
by
\begin{align}\label{iso map}
T(s(x))=s(\lambda x), \forall s(x)\in \mathbb{Z}_{p}[u,v][x]/\langle x^{p-1}-1\rangle.
\end{align}

\begin{theorem}\label{iso th}
The map $T$ defined in equation (\ref{iso map}) is a ring isomorphism.
\end{theorem}

\begin{proof}
Let $s_{1}(x)\equiv s_{2}(x)$ mod $(x^{p-1}-1)$. Then $s_{1}(x)-s_{2}(x)=k(x) (x^{p-1}-1)$ that is, $ s_{1}(\lambda x)-s_{2}(\lambda x)=k(\lambda x)(\lambda^{p-1} x^{p-1}-1)$, implies $ s_{1}(\lambda x)-s_{2}(\lambda x)=\lambda^{p-1} k(\lambda x) (x^{p-1}-\lambda)$. Hence, $s_{1}(\lambda x)\equiv s_{2}(\lambda x)$ mod $(x^{p-1}-\lambda)$. Rest parts are easy to verify.
\end{proof}

\begin{corollary}\label{iso cor}
A linear code $\mathcal{C}$ of length $(p-1)$ is cyclic over $\mathbb{Z}_{p}[u,v]$ if and only if $T(\mathcal{C})$ is a $\lambda$-constacyclic code over $\mathbb{Z}_{p}[u,v]$.
\end{corollary}

\begin{proof}
Simple consequence of Theorem \ref{iso th}.
\end{proof}
By Corollary \ref{iso cor} and part (2) of Theorem \ref{gen th 1}, we characterize  $\lambda$ -constacyclic code over $\mathbb{Z}_{p}[u,v]$ as in the next theorem.

\begin{theorem}\label{th cons gen}
Let $\mathcal{C}$ be a $\lambda$-constacyclic code of length $(p-1)$ over $\mathbb{Z}_{p}[u,v]$. Then $\mathcal{C}$ is an ideal of $\mathbb{Z}_{p}[u,v][x]/\langle x^{p-1}-\lambda \rangle$ given by
\begin{align*}
\mathcal{C}=\langle g(\lambda x)+ua(\lambda x)+vp_{1}(\lambda x), vb(\lambda x)\rangle,
\end{align*}
where $a\mid g\mid (x^{p-1}-1), b\mid g\mid (x^{p-1}-1)$ and $b\mid p_{1}(\frac{x^{p-1}-1}{a})$.
\end{theorem}

\section{Structure of $\mathbb{Z}_{p}\mathbb{Z}_{p}[u,v]$-additive constacyclic codes}\label{sec7}

\begin{definition}
Any non empty subset $\mathcal{C}$ of $\mathbb{Z}_{p}^{\alpha}\times \mathbb{Z}_{p}[u,v]^{\beta}$ is said to be a $\mathbb{Z}_{p}\mathbb{Z}_{p}[u,v]$-additive $\lambda$-constacyclic code of length $(\alpha, \beta)$ if
\begin{enumerate}
\item $\mathcal{C}$ is $\mathbb{Z}_{p}\mathbb{Z}_{p}[u,v]$-additive, and
\item for any $z=(c,d)=(c_{0},c_{1},\cdots,c_{\alpha-1},d_{0},d_{1},\cdots,d_{\beta-1})\in \mathcal{C}$, we have\\
$\tau_{\lambda}(z)=(c_{\alpha-1},c_{0},\cdots,c_{\alpha-2},\lambda d_{\beta-1},d_{0},\cdots,d_{\beta-2})\in \mathcal{C}$.
\end{enumerate}
\end{definition}
To consider $R_{\alpha,\beta,\lambda}=\mathbb{Z}_{p}[x]/\langle x^{\beta}-1\rangle\times \mathbb{Z}_{p}[u,v][x]/\langle x^{\beta}-\lambda\rangle$ as a module over $\mathbb{Z}_{p}[u,v][x]$, we use the multiplication $\ast$
defined in preliminary as scalar multiplication. In fact, the set $R_{\alpha,\beta,\lambda}=\mathbb{Z}_{p}[x]/\langle x^{\beta}-1\rangle\times \mathbb{Z}_{p}[u,v][x]/\langle x^{\beta}-\lambda\rangle$ is a $\mathbb{Z}_{p}[u,v][x]$-module under the multiplication $\ast$
as defined in preliminary.

\begin{theorem}
Let $\mathcal{C}$ be a $\mathbb{Z}_{p}\mathbb{Z}_{p}[u,v]$-additive code of length $(\alpha, \beta)$. Then $\mathcal{C}$ is a $\mathbb{Z}_{p}\mathbb{Z}_{p}[u,v]$-additive $\lambda$-constacyclic code if and only if $\mathcal{C}$ is a $\mathbb{Z}_{p}[u,v][x]$-submodule of $R_{\alpha,\beta,\lambda}$.
\end{theorem}
\begin{proof}
Same as the proof of Theorem \ref{th basic}.
\end{proof}

To find the unique set of generators for $\mathbb{Z}_{p}\mathbb{Z}_{p}[u,v]$-additive $\lambda$-constacyclic code, we follow the same technique as used in case of $\mathbb{Z}_{p}\mathbb{Z}_{p}[u,v]$-additive cyclic codes in Section \ref{sec4}. Therefore, for any prime $p>2$,
we define the projection map
\begin{align*}
\Omega: \mathbb{Z}_{p}[x]/\langle x^{\alpha}-1\rangle\times \mathbb{Z}_{p}[u,v][x]/\langle x^{p-1}-\lambda\rangle \longrightarrow \mathbb{Z}_{p}[u,v][x]/\langle x^{p-1}-\lambda\rangle
\end{align*}
by
\begin{align*}
\Omega(c(x),d(x))=d(x).
\end{align*}
Let $\mathcal{C}$ be a $\mathbb{Z}_{p}\mathbb{Z}_{p}[u,v]$-additive $\lambda$-constacyclic code of length $(\alpha, p-1)$ and consider the restriction on $\Omega$ to $\mathcal{C}$. Then $\Omega$ is a $\mathbb{Z}_{p}[u,v][x]$-module homomorphism with $ker(\Omega|_\mathcal{C})=\langle (f_{1},0)\rangle$ where $f_{1}\mid (x^{\alpha}-1)$.
Also, the homomorphic image of $\mathcal{C}$, denoted by $\Omega(\mathcal{C})$, is an ideal of $\mathbb{Z}_{p}[u,v][x]/\langle x^{p-1}-\lambda\rangle$,
therefore $\Omega(\mathcal{C})$ is a $\lambda$-constacyclic code over $\mathbb{Z}_{p}[u,v]$ of length $(p-1)$. Hence, with the help of Theorem \ref{th cons gen} and
of the above discussion, we conclude the next Theorem as follows.

\begin{theorem}\label{th gen consta}
Let $\mathcal{C}$ be a $\mathbb{Z}_{p}\mathbb{Z}_{p}[u,v]$-additive $\lambda$-constacyclic code of length $(\alpha, p-1)$. Then $\mathcal{C}$ is a $\mathbb{Z}_{p}[u,v][x]$-submodule of $R_{\alpha,\beta,\lambda}$ given by
\begin{align*}
\mathcal{C}=\langle (f_{1}(x),0),(f_{2}(x),g(\lambda x)+ua(\lambda x)+vp_{1}(\lambda x)),(f_{3}(x),vb(\lambda x))\rangle,
\end{align*}
where $f_{1}\mid (x^{\alpha}-1), a\mid g\mid (x^{p-1}-1), b\mid g\mid (x^{p-1}-1)$ and $f_{i}\in \mathbb{Z}_{p}[x]$, for $i=2,3.$
\end{theorem}

\begin{theorem}\label{th min consta}
Let $\mathcal{C}$ be a $\mathbb{Z}_{p}\mathbb{Z}_{p}[u,v]$-additive $\lambda$-constacyclic code of length $(\alpha, p-1)$ given by
\begin{align*}
   \mathcal{C}=\langle (f_{1}(x),0),(f_{2}(x),g(\lambda x)+ua(\lambda x)+vp_{1}(\lambda x)),(f_{3}(x),vb(\lambda x))\rangle,
\end{align*}{}
where $f_{1}\mid (x^{\alpha}-1), a\mid g\mid (x^{p-1}-1)$, $ b\mid g\mid (x^{p-1}-1)$ and $f_{i}\in \mathbb{Z}_{p}[x]$, for $i=2,3$. Let $h=\frac{x^{p-1}-1}{g}, m_{1}=gcd(ha, hp_{1}, x^{p-1}-1), m_{2}=\frac{x^{p-1}-1}{m_{1}}, deg(f_{1})=t_{1}, deg(g)=t_{2}, deg(m_{2})=t_{3}, deg(b)=t_{4}$ and
\begin{align*}
S_{1}&=\bigcup_{i=0}^{\alpha-t_{1}-1}\{ x^{i}\ast(f_{1}(x),0) \};\\
S_{2}&=\bigcup_{i=0}^{p-t_{2}-2}\{ x^{i}\ast(f_{2}(x),g(\lambda x)+ua(\lambda x)+vp_{1}(\lambda x) ) \};\\
S_{3}&=\bigcup_{i=0}^{p-t_{3}-2}\{ x^{i}\ast(h(x)f_{2}(x),uh(x)a(\lambda x)+vh(x)p_{1}(\lambda x) ) \};\\
S_{4}&=\bigcup_{i=0}^{p-t_{4}-2}\{ x^{i}\ast(f_{3}(x),vb(\lambda x) ) \}.
\end{align*}
Then $S=S_{1}\cup S_{2}\cup S_{3}\cup S_{4}$ is a minimal spanning set for the code $\mathcal{C}.$  Moreover, $\mid \mathcal{C}\mid=p^{\alpha+5p-(t_{1}+t_{2}+t_{3}+t_{4}+5)}$.
\end{theorem}

\begin{proof}
Same as the proof of Theorem \ref{ad gn th 1}.
\end{proof}

\begin{corollary}
Let $\mathcal{C}$ be a $\mathbb{Z}_{p}\mathbb{Z}_{p}[u,v]$-additive $\lambda$-constacyclic code of length $(\alpha, p-1)$ where
\begin{align*}
    \mathcal{C}=\langle (f_{1}(x),0),(f_{2}(x),g(\lambda x)+ua(\lambda x)+vp_{1}(\lambda x)),(f_{3}(x),vb(\lambda x))\rangle,
\end{align*}{}
as given in Theorem \ref{th gen consta}. Then any codeword $c(x)\in \mathcal{C}$ is of the form
\begin{align*}
c(x)=&s_{1}(x)\ast (f_{1}(x),0)+s_{2}(x)\ast (f_{2}(x),g(\lambda x)+ua(\lambda x)+vp_{1}(\lambda x))+\\
&s_{3}(x)\ast (h(x)f_{2}(x),uh(x)a(\lambda x)+vh(x)p_{1}(\lambda x) )+s_{4}(x)\ast (f_{3}(x),vb(\lambda x)),
\end{align*}
where $s_{1},s_{3},s_{4}\in \mathbb{Z}_{p}[x]$ with $deg(s_{1})\leq (\alpha-t_{1}-1) $  and  $s_{2}\in \mathbb{Z}_{p}[u,v][x]$ with $deg(s_{i})\leq (p-t_{i}-2)$ for $i=2,3,4.$
\end{corollary}
\begin{proof}
It is a consequences of Theorem \ref{th min consta}.
\end{proof}

Let $\mathcal{C}$ be a $\mathbb{Z}_{p}\mathbb{Z}_{p}[u,v]$-additive code of length $(\alpha,\beta)$. Let $P_{\alpha}$ be the projection which maps each codeword
into the first $\alpha$ co-ordinates, and let $P_{\beta}$ be the projection which maps each codeword into the last $\beta$ co-ordinates. Since, projection maps are linear maps, so $\mathcal{C}_{\alpha}=P_{\alpha}(\mathcal{C})$ is a linear code of length $\alpha$ over $\mathbb{Z}_{p}$ and $\mathcal{C}_{\beta}=P_{\beta}(\mathcal{C})$ is a linear code of length $\beta$ over $\mathbb{Z}_{p}[u,v]$. If $\mathcal{C}=\mathcal{C}_{\alpha}\times \mathcal{C}_{\beta}$, then $\mathcal{C}$ is said to be separable.

\begin{theorem}\label{th sepa}
Let $\mathcal{C}=\mathcal{C}_{\alpha}\times \mathcal{C}_{\beta}$ be a $\mathbb{Z}_{p}\mathbb{Z}_{p}[u,v]$-additive code of length $(\alpha,\beta)$ where $\mathcal{C}_{\alpha}$ is a linear code of length $\alpha$ over $\mathbb{Z}_{p}$ and $\mathcal{C}_{\beta}$ is a linear code of length $\beta$ over $\mathbb{Z}_{p}[u,v]$. Then $\mathcal{C}$ is $\mathbb{Z}_{p}\mathbb{Z}_{p}[u,v]$-additive $\lambda$-constacyclic code if and only if $\mathcal{C}_{\alpha}$ is cyclic and $\mathcal{C}_{\beta}$ is $\lambda$-constacyclic code.
\end{theorem}

\begin{proof}
Let $\mathcal{C}=\mathcal{C}_{\alpha}\times \mathcal{C}_{\beta}$ be a $\mathbb{Z}_{p}\mathbb{Z}_{p}[u,v]$-additive $\lambda$-constacyclic code. Let $(c_{0},c_{1},\cdots,c_{\alpha-1})\in \mathcal{C}_{\alpha}$ and $(d_{0},d_{1},\cdots,d_{\beta-1})\in \mathcal{C}_{\beta}$. Therefore, $z=(c_{0},c_{1},\cdots,c_{\alpha-1},d_{0},d_{1},\dots,d_{\beta-1})\in \mathcal{C}$. Since, $\mathcal{C}$ is $\mathbb{Z}_{p}\mathbb{Z}_{p}[u,v]$-additive $\lambda$-constacyclic code, so  $\tau_{\lambda}(z)=(c_{\alpha-1},c_{0},\cdots,c_{\alpha-2},\lambda d_{\beta-1},d_{0},\cdots,d_{\beta-2})\in \mathcal{C}$. Hence, $(c_{\alpha-1},c_{0},\cdots,c_{\alpha-2})\in \mathcal{C}_{\alpha}$ and $(\lambda d_{\beta-1},d_{0},\cdots,d_{\beta-2})\in \mathcal{C}_{\beta}$. This implies $\mathcal{C}_{\alpha}$ is a cyclic and $\mathcal{C}_{\beta}$ is a $\lambda$-constacyclic code.

Conversely, let $\mathcal{C}_{\alpha}$ be cyclic, $\mathcal{C}_{\beta}$ be a $\lambda$-constacyclic code and $(c_{0},c_{1},\cdots,c_{\alpha-1},d_{0},d_{1},\cdots,\\d_{\beta-1})\in \mathcal{C}$. Therefore, $(c_{0},c_{1},\cdots,c_{\alpha-1})\in \mathcal{C}_{\alpha}$ and $( d_{0},d_{1},\cdots,d_{\beta-1})\in \mathcal{C}_{\beta}$ and hence, $(c_{\alpha-1},c_{0},\\\cdots,c_{\alpha-2})\in \mathcal{C}_{\alpha}$ and $(\lambda d_{\beta-1},d_{0},\cdots,d_{\beta-2})\in \mathcal{C}_{\beta}$. Therefore, $\tau_{\lambda}(z)=(c_{\alpha-1},c_{0},\cdots,c_{\alpha-2},\lambda d_{\beta-1},\\d_{0},\cdots,d_{\beta-2})\in \mathcal{C}$. Thus, $\mathcal{C}$ is a $\mathbb{Z}_{p}\mathbb{Z}_{p}[u,v]$-additive $\lambda$-constacyclic code.
\end{proof}

\begin{corollary}
 Let $\mathcal{C}=\mathcal{C}_{\alpha}\times \mathcal{C}_{\beta}$ be a $\mathbb{Z}_{p}\mathbb{Z}_{p}[u,v]$-additive code of length $(\alpha,\beta)$ where $\mathcal{C}_{\alpha}$ is a linear code of length $\alpha$ over $\mathbb{Z}_{p}$ and  $\mathcal{C}_{\beta}$ is a linear code of length $\beta$ over $\mathbb{Z}_{p}[u,v]$. Then $\mathcal{C}$ is a $\mathbb{Z}_{p}\mathbb{Z}_{p}[u,v]$-additive cyclic code if and only if $\mathcal{C}_{\alpha}$ and $\mathcal{C}_{\beta}$ both are cyclic codes.
\end{corollary}
\begin{proof}
A particular case of Theorem \ref{th sepa} with $\lambda=1$.
\end{proof}

\section{Gray maps and $\mathbb{Z}_p$-images of $\mathbb{Z}_p\mathbb{Z}_p[u,v]$-additive constacyclic codes}\label{sec8}
We define two maps $\phi_1,\phi_2: \mathbb{Z}_p[u,v]\longrightarrow \mathbb{Z}_p^2$ by $\phi_1(a+ub+vc)=(b+c-2a,b+c)$ and $\phi_2(a+ub+vc)=(b-c,a-b+c)$ where $a,b,c\in \mathbb{Z}_p$.
The maps $\phi_{1},\phi_{2}$ are linear and their extensions $\phi_{1},\phi_{2}: \mathbb{Z}_{p}[u,v]^{\beta}\longrightarrow \mathbb{Z}_{p}^{2\beta}$ are respectively defined by
\begin{align}\label{Gray 1}
\nonumber \phi_{1}(r_{0},r_{1},\cdots,r_{\beta-1})=&(b_0+c_0-2a_{0},b_1+c_1-2a_{1},\cdots,b_{\beta-1}+c_{\beta-1}-2a_{\beta-1},\\
&b_{0}+c_0,b_{1}+c_1,\cdots,b_{\beta-1}+c_{\beta-1}),
\end{align}
and
\begin{align}\label{Gray 2}
\nonumber \phi_{2}(r_{0},r_{1},\cdots,r_{\beta-1})=&(b_{0}-c_0,b_{1}-c_1,\cdots,b_{\beta-1}-c_{\beta-1},a_0-b_{0}+c_0,\\
&a_1-b_{1}+c_1,\cdots,a_{\beta-1}-b_{\beta-1}+c_{\beta-1}),
\end{align}
where $r_{i}=a_{i}+ub_{i}+vc_i\in \mathbb{Z}_{p}[u,v]$ for $i=0,1,\cdots,\beta-1.$ These maps are isometric and the Gray distance is defined as $d_{G}(r,r')=w_{G}(r-r'),$
where $w_{G}(r)=w_{H}(\phi_{1}(r)),w_{G}(r)=w_{H}(\phi_{2}(r))$, respectively for $r,r'\in \mathbb{Z}_{p}[u,v]^{\beta}$ and $w_H$ denotes the Hamming weight in $\mathbb{Z}_p$.

\begin{lemma}\label{lemma consta 1}
Let $\phi_{1}$ be the Gray map defined in equation (\ref{Gray 1}), $\Upsilon_{1+u+v},$ the $(1+u+v)$-constacyclic shift and $\sigma,$ the cyclic shift. Then $\phi_{1}\Upsilon_{1+u+v}=\sigma\phi_{1}$.
\end{lemma}

\begin{proof}
Let $r_{i}=a_{i}+ub_{i}+vc_i\in \mathbb{Z}_{p}[u,v]$ for $i=0,1,\cdots,\beta-1$ and $r=(r_{0},r_{1},\cdots,r_{\beta-1})\in \mathbb{Z}_{p}[u,v]^{\beta}$. Now,
\begin{align*}
\phi_{1}\Upsilon_{1+u+v}(r)=&\phi_{1}((1+u+v) r_{\beta-1}, r_{0},\dots,r_{\beta-2})\\
=&(b_{\beta-1}+c_{\beta-1},b_{0}+c_{0}-2a_{0},\cdots,b_{\beta-2}+c_{\beta-2}-2a_{\beta-2},\\
&b_{\beta-1}+c_{\beta-1}-2a_{\beta-1},b_{0}+c_0,b_{1}+c_1,\cdots,b_{\beta-2}+c_{\beta-2}).
\end{align*}
Also, we have
\begin{align*}
\sigma\phi_{1}(r)=&\sigma(b_{0}+c_{0}-2a_{0},\cdots,b_{\beta-2}+c_{\beta-2}-2a_{\beta-2},b_{\beta-1}+c_{\beta-1}-2a_{\beta-1},\\
&b_{0}+c_0,b_{1}+c_1,\cdots,b_{\beta-2}+c_{\beta-2},b_{\beta-1}+c_{\beta-1})\\
=&(b_{\beta-1}+c_{\beta-1},b_{0}+c_{0}-2a_{0},\cdots,b_{\beta-2}+c_{\beta-2}-2a_{\beta-2},\\
&b_{\beta-1}+c_{\beta-1}-2a_{\beta-1},b_{0}+c_0,b_{1}+c_1,\cdots,b_{\beta-2}+c_{\beta-2}).
\end{align*}
Therefore, $\phi_{1}\Upsilon_{1+u+v}=\sigma\phi_{1}$.
\end{proof}

\begin{theorem}
Let $\mathcal{C}$ be a $(1+u+v)$-constacyclic code of length $\beta$ over $\mathbb{Z}_{p}[u,v]$. Then $\phi_{1}(\mathcal{C})$ is a cyclic code of length $2\beta$ over $\mathbb{Z}_{p}$.
\end{theorem}

\begin{proof}
Let $\mathcal{C}$ be a $(1+u+v)$-constacyclic code. Then $\Upsilon_{1+u+v}(\mathcal{C})=\mathcal{C}$. Also, by Lemma \ref{lemma consta 1}, we have $\phi_{1}\Upsilon_{1+u+v}(\mathcal{C})=\phi_{1}(\mathcal{C})=\sigma(\phi_{1}(\mathcal{C}))$. This shows that $\phi_{1}(\mathcal{C})$ is a cyclic code of length $2\beta$ over $\mathbb{Z}_{p}$.
\end{proof}

\begin{lemma}\label{lemma consta 2}
Let $\phi_{2}$ be the Gray map defined in equation (\ref{Gray 2}), $\Upsilon_{1+u+v}$ be the $(1+u+v)$-constacyclic shift and $\pi_2$ be the quasi-cyclic shift operator defined in equation (\ref{map quasi-cyclic}). Then $\phi_{2}\Upsilon_{1+u+v}=\pi_2\phi_{2}$.
\end{lemma}
\begin{proof}
Let $r_{i}=a_{i}+ub_{i}+vc_i\in \mathbb{Z}_{p}[u,v]$ for $i=0,1,\cdots,\beta-1$ and $r=(r_{0},r_{1},\cdots,r_{\beta-1})\in \mathbb{Z}_{p}[u,v]^{\beta}$. Now,
\begin{align*}
\phi_{2}\Upsilon_{1+u+v}(r)=&\phi_{2}((1+u+v) r_{\beta-1},r_{0},\cdots,r_{\beta-2})\\
=&(b_{\beta-1}-c_{\beta-1},b_{0}-c_0,\cdots,b_{\beta-2}-c_{\beta-2},a_{\beta-1}-b_{\beta-1}+c_{\beta-1},\\
&a_0-b_{0}+c_0,\cdots,a_{\beta-2}-b_{\beta-2}+c_{\beta-2}).
\end{align*}
On the other hand,
\begin{align*}
\pi_2\phi_{2}(r)=&\pi_2(b_{0}-c_0,\cdots,b_{\beta-2}-c_{\beta-2},b_{\beta-1}-c_{\beta-1},a_0-b_{0}+c_0,\cdots,\\
&a_{\beta-2}-b_{\beta-2}+c_{\beta-2},a_{\beta-1}-b_{\beta-1}+c_{\beta-1})\\
=&(b_{\beta-1}-c_{\beta-1},b_{0}-c_0,\cdots,b_{\beta-2}-c_{\beta-2},a_{\beta-1}-b_{\beta-1}+c_{\beta-1},\\
&a_0-b_{0}+c_0,\cdots,a_{\beta-2}-b_{\beta-2}+c_{\beta-2}).
\end{align*}
Hence, $\phi_{2}\Upsilon_{1+u+v}=\pi_2\phi_{2}$.
\end{proof}
\begin{theorem}
Let $\mathcal{C}$ be a $(1+u+v)$-constacyclic code of length $\beta$ over $\mathbb{Z}_{p}[u,v]$. Then $\phi_{2}(\mathcal{C})$ is a quasi-cyclic code of length $2 \beta$ and index $2$ over $\mathbb{Z}_{p}.$
\end{theorem}
\begin{proof}
Since $\mathcal{C}$ is a $(1+u+v)$-constacyclic code, so $\Upsilon_{1+u+v}(\mathcal{C})=\mathcal{C}$. By Lemma \ref{lemma consta 2}, we have $\phi_{2}\Upsilon_{1+u+v}(\mathcal{C})=\phi_{2}(\mathcal{C})=\pi_{2}(\phi_{2}(\mathcal{C}))$. Therefore, $\phi_{2}(\mathcal{C})$ is a quasi-cyclic code of length $2\beta$ and index $2$ over $\mathbb{Z}_{p}.$
\end{proof}

Now, we define the maps $\Phi_{1},\Phi_{2}: \mathbb{Z}_{p}\times \mathbb{Z}_{p}[u,v]\longrightarrow \mathbb{Z}_{p}^{3}$ with the help of the map $\phi_{1},\phi_{2}$, respectively by
\begin{align}\label{map big 1}
\Phi_{1}(e, a+ub+vc)=(e,\phi_{1}(a+ub+vc))=(e,b+c-2a,b+c), ~~a, b,c \in \mathbb{Z}_{p},
\end{align}
and
\begin{align}\label{map big 2}
\Phi_{2}(e, a+ub+vc)=(e,\phi_{2}(a+ub+vc))=(e,b-c,a-b+c), ~~a, b,c \in \mathbb{Z}_{p}.
\end{align}
Again, these maps can be extended from $\mathbb{Z}_{4}^{\alpha}\times \mathbb{Z}_{4}[u]^{\beta}\longrightarrow \mathbb{Z}_{4}^{\alpha+2\beta}$ like maps $\phi_{1},\phi_{2}$.

\begin{theorem}\label{thm qc}
Let $\mathcal{C}$ be a $\mathbb{Z}_{p}\mathbb{Z}_{p}[u,v]$-additive $(1+u+v)$-constacyclic code of length $(\alpha,\beta)$. Then $\Phi_{1}(\mathcal{C})$ is a generalized QC code of block length $(\alpha, 2\beta)$ over $\mathbb{Z}_{p}$.
\end{theorem}

\begin{proof}
Let $(e_{0},e_{1},\cdots,e_{\alpha-1},b_0+c_0-2a_{0},b_1+c_1-2a_{1},\cdots,b_{\beta-1}+c_{\beta-1}-2a_{\beta-1},b_{0}+c_0,b_{1}+c_1,\cdots,b_{\beta-1}+c_{\beta-1})\in \Phi_{1}(\mathcal{C})$.\\ Therefore, there exists $(e_{0},e_{1},\cdots,e_{\alpha-1},r_{0},r_{1},\cdots,r_{\beta-1})\in \mathcal{C}$ such that
\begin{align*}
&\Phi_{1}(e_{0},e_{1},\cdots,e_{\alpha-1},r_{0},r_{1},\cdots,r_{\beta-1})\\
=&(e_{0},e_{1},\cdots,e_{\alpha-1},b_0+c_0-2a_{0},b_1+c_1-2a_{1},\cdots,b_{\beta-1}+c_{\beta-1}-2a_{\beta-1},\\
&b_{0}+c_0,b_{1}+c_1,\cdots,b_{\beta-1}+c_{\beta-1})\in \Phi_{1}(\mathcal{C})
\end{align*}
where $r_{i}=a_{i}+ub_{i}+vc_i\in \mathbb{Z}_{p}[u,v]$ and $a_{i},b_{i},c_{i}\in \mathbb{Z}_{p}$ for $i=0,1,\cdots,\beta-1.$
Since $\mathcal{C}$ is a $\mathbb{Z}_{p}\mathbb{Z}_{p}[u,v]$-additive $(1+u+v)$-constacyclic code, $(e_{\alpha-1},e_{0},\cdots,e_{\alpha-2},(1+u+v) r_{\beta-1},r_{0},\cdots,r_{\beta-2})\in \mathcal{C}$. Therefore,
\begin{align*}
&\Phi_{1}(e_{\alpha-1},e_{0},\cdots,e_{\alpha-2},(1+u+v) r_{\beta-1},r_{0},\cdots,r_{\beta-2})\\
=&(e_{\alpha-1},e_{0},\cdots,e_{\alpha-2},b_{\beta-1}+c_{\beta-1}, b_0+c_0-2a_{0},b_1+c_1-2a_{1},\cdots,b_{\beta-1}+\\
&c_{\beta-1}-2a_{\beta-1},b_{0}+c_0,b_{1}+c_1,\cdots,b_{\beta-2}+c_{\beta-2}).
\end{align*}
This shows that $\Phi_{1}(\mathcal{C})$ is a generalized QC code of block length $(\alpha, 2\beta)$ over $\mathbb{Z}_{p}$.
\end{proof}
\begin{corollary}\label{cor qc}
 If $\mathcal{C}$ is a $\mathbb{Z}_{p}\mathbb{Z}_{p}[u,v]$-additive $(1+u+v)$-constacyclic code of length $(2\alpha,\alpha)$, then $\Phi_1(\mathcal{C})$ is a quasi-cyclic code of length $3\alpha$ and index $3$ over $\mathbb{Z}_{p}$.
\end{corollary}{}

\begin{proof}
It follows from  Theorem \ref{thm qc}, for $\alpha=2\beta$.
\end{proof}

\begin{theorem}\label{thm qc2}
Let $\mathcal{C}$ be a $\mathbb{Z}_{p}\mathbb{Z}_{p}[u,v]$-additive $(1+u+v)$-constacyclic code of length $(\alpha,\beta)$. Then $\Phi_{2}(\mathcal{C})$ is a generalized QC code of block length $(\alpha,\beta,\beta)$ over $\mathbb{Z}_{p}$.
\end{theorem}
\begin{proof}
Same as the proof of  Theorem \ref{thm qc}.
\end{proof}

\begin{corollary}\label{cor qc2}
 If $\mathcal{C}$ is a $\mathbb{Z}_{p}\mathbb{Z}_{p}[u,v]$-additive $(1+u+v)$-constacyclic code of length $(\alpha,\alpha)$, then $\Phi_2(\mathcal{C})$ is a quasi-cyclic code of length $3\alpha$ and index $3$ over $\mathbb{Z}_{p}$.
\end{corollary}

\section{Examples}\label{sec9}
In this section, we present several optimal and new codes from these class of constacyclic codes. From  Theorem \ref{th cons gen}, we recall
that the $(1+u+v)$-constacyclic code $\mathcal{C}$ of length $(p-1)$ over $\mathbb{Z}_p[u,v]$ is given by
\begin{align*}
\mathcal{C}=\langle g(x+ux+vx)+ua(x+ux+vx)+vp_{1}(x+ux+vx), vb(x+ux+vx)\rangle,
\end{align*}
where $a\mid g\mid (x^{p-1}-1), b\mid g\mid (x^{p-1}-1)$ and $b\mid p_{1}(\frac{x^{p-1}-1}{a})$.\\ For brevity of notation,
we can write $\mathcal{C}=\langle h(x),k(x)\rangle$, where $h(x)=g(x+ux+vx)+ua(x+ux+vx)+vp_{1}(x+ux+vx), k(x)=vb(x+ux+vx)$.
In Table \ref{Tab1}, the second and third column represent the generator polynomials $h(x),k(x)$ respectively.  We write coefficients
of generator polynomials in decreasing order, for example, we write $[1,5+5u+5v,0,1+2u+2v,5+3v]$ to represent the polynomial $x^4+(5+5u+5v)x^3+(1+2u+2v)x+5+3v$.
The fourth and fifth columns give the $\phi_{1}$-Gray and $\phi_{2}$-Gray images of the constacyclic codes, respectively.
The codes marked by the symbol $*$ are optimal according to the database \cite{Grassl}. Till now, there is no database available for
the linear codes over the fields $GF(q)$, $q>9$, and as per the available literature \cite{Aydin19,Chen15}, we found that in Table \ref{Tab1}, the linear codes
marked by the symbol $\#$ are the good and new codes over the fields $\mathbb{Z}_{11},\mathbb{Z}_{13}$ and $\mathbb{Z}_{17}$.

\begin{landscape}
\begin{table}[ht!]
\caption{Gray images of $(1+u+v)$-constacyclic codes over $\mathbb Z_p[u,v]$ of length $(p-1)$}\label{Tab1}
\vspace{0.5cm}
\renewcommand{\arraystretch}{1.8}
\begin{center}
\begin{tabular}{|c|c|c|c|c|}
\hline
 $p$ & $h(x)$ & $k(x)$  & $\phi_{1}(\mathcal{C})$ & $\phi_{2}(\mathcal{C})$\\
\hline
 %$5$ & $[1,4+5u+5v,3+u+v]$ & $[v,3v]$ & $[8,6,2]^*_{5}$ & $[8,8,1]^*_{5}$ \\
%\hline
 $5$ & $[1+u+v,2,4,3+2u+v]$ & $[v,4v]$ & $[8,4,4]^*_{5}$ & $[8,4,4]^*_{5}$ \\
\hline
 $5$ & $[1,4+4v,3+u+v]$ & $[v,3v]$ & $[8,5,3]^*_{5}$ & $[8,5,3]^*_{5}$ \\
\hline
 $7$ & $[1,5+5u+5v,0,1+2u+2v,5+3v]$ & $[v,4v]$ & $[12,7,4]_{7}$ & $[12,7,4]_{7}$ \\
\hline
 $7$ & $[1+u+v,2,5+6u+6v,3+3u+v]$ & $[v,2v]$ & $[12,8,3]_{7}$ & $[12,8,3]_{7}$ \\
\hline
 $7$ & $[1,1+2u+2v,1+3u+v]$ & $[v,5v]$ & $[12,9,3]^*_{7}$ & $[12,9,3]^*_{7}$ \\
\hline

 $11$ & $[1,8+9u+8v,4+3u+v]$ & $[v,5v]$ & $[20,17,2]^{\#}_{11}$ & $[20,17,3]^{\#}_{11}$ \\
\hline
 $11$ & $[1+u+v,1,u,4+2u+v]$ & $[v,6v]$ & $[20,16,3]^{\#}_{11}$ & $[20,16,3]^{\#}_{11}$ \\
\hline

 $11$ & $[1+u+v,6,6+6u+6v,10,9+9u+9v,10+u,9+6u+10v,10+4u+v]$ & $[v,5v,8v]$ & $[20,12,4]^{\#}_{11}$ & $[20,12,4]^{\#}_{11}$ \\
\hline
 $13$ & $[1+u+v,7,9+10u+10v,11+2u+v]$ & $[v,11v]$ & $[24,20,3]^{\#}_{13}$ & $[24,20,2]^{\#}_{13}$ \\
\hline
 $17$ & $[1+u+v,4,7+8u+7v,6+2u+v]$ & $[v,8v]$ & $[34,30,3]^{\#}_{17}$ & $[34,30,3]^{\#}_{17}$ \\
\hline
\end{tabular}
\end{center}
\end{table}
\end{landscape}

\begin{example}
Let $p=5$, and $\mathcal{C}$ be a $\mathbb{Z}_{5}\mathbb{Z}_{5}[u,v]$-additive $(1+u+v)$-constacyclic code of length $(3,4)$. Then $\mathcal{C}$ is a $\mathbb{Z}_{5}\mathbb{Z}_{5}[u,v][x]$-submodule of $R_{3,4,1+u+v}=\mathbb{Z}_{5}[x]/\langle x^{3}-1\rangle\times \mathbb{Z}_{5}[u,v][x]/\langle x^{4}-(1+u+v)\rangle$ given by Theorem \ref{th gen consta}. Take $f_1(x)=x+4, g(x)=x+3=a(x),b(x)=1,p_1(x)=1,f_2(x)=f_3(x)=x+1$. Then, we have $\mathcal{C}=\langle (x+4,0),(x+1,(1+2u+v)x+3+3u+v),(x+1,v)\rangle$. Therefore, $\Phi_1(\mathcal{C}),\Phi_2(\mathcal{C})$ have the generator matrices
\[
\left[ {\begin{array}{ccccccccccc}
1 &0 &0 &0 &0 &0 &0 &0 &0 &4 &3\\
0 &1 &0 &0 &0 &0 &0 &0 &0 &4 &3\\
0 &0 &1 &0 &0 &0 &0 &0 &0 &4 &3\\
0 &0 &0 &1 &0 &0 &0 &0 &0 &0 &2\\
0 &0 &0 &0 &1 &0 &0 &0 &0 &1 &1\\
0 &0 &0 &0 &0 &1 &0 &0 &0 &3 &4\\
0 &0 &0 &0 &0 &0 &1 &0 &0 &2 &0\\
0 &0& 0 &0 &0 &0& 0 &1 &0 &2 &2\\
0 &0 &0 &0 &0& 0 &0 &0& 1 &1 &3
\end{array} } \right],
\left[ {\begin{array}{ccccccccccc}
1 &0 &0 &0 &0 &0 &0 &0 &0 &1 &2\\
0 &1 &0 &0 &0 &0 &0 &0 &0 &1 &2\\
0 &0 &1 & 0&0 &0& 0 &0 &0 & 1&2\\
0 &0 &0 &1 &0 &0 &0 &0 &0 &4& 1\\
0 &0 &0 &0 &1 &0 &0 &0 &0 &3 &2\\
0 &0 &0 &0 &0 &1 &0 &0 &0 &1 &4\\
0 &0 &0 &0 &0 &0 &1 &0 &0& 2 &3\\
0 &0 &0 &0& 0 &0 &0 &1& 0 &2 &2\\
0 &0 &0& 0& 0& 0& 0 &0& 1 &1& 3
\end{array} } \right]
\]
respectively, and both have the parameters $[11,9,2]$ over $\mathbb{Z}_5$, which is an optimal code according to database \cite{Grassl}.
\end{example}

\begin{example}
Let $p=7,\alpha=4$ and $\mathcal{C}$ be a $\mathbb{Z}_{7}\mathbb{Z}_{7}[u,v]$-additive $(1+u+v)$-constacyclic code of length $(\alpha,p-1)=(4,6)$. Further, let $g(x)=(x+2)(x+3)(x+4),a(x)=x+4,b(x)=x^2+5x+6,f_1(x)=x+6,p_1(x)=1$. Then, the consatcyclic code is given by $\mathcal{C}=\langle (x+6,0),(x+1,(1+u+v)x^3+2x^2+(5+6u+5v)x+3+4u+v),(x+1,vx^2+5vx+6v)\rangle$.
Moreover, $\Phi_1(\mathcal{C})$ and $\Phi_2(\mathcal{C})$ are linear codes over $\mathbb{Z}_7$ with parameters $[16,12,2]$.

\end{example}

\section{Conclusion and future work}\label{sec10}
In this article, we have derived the generator polynomials of cyclic and constacyclic codes over $\mathbb{Z}_{p}[u,v]$ and also calculated
the minimal generating set for cyclic codes of any length $\beta.$ Further, we identify additive cyclic, additive $\lambda$-constacyclic codes as
the $\mathbb{Z}_{p}[u,v][x]$-submodules of $R_{\alpha,\beta}$ and $R_{\alpha,\beta,\lambda},$ respectively. Moreover, we have calculated the
generator polynomials as well as the minimal spanning sets for additive cyclic codes of length $(\alpha,\beta),$ and for additive $\lambda$-constacyclic
codes with length $(\alpha,p-1)$. Finally, some optimal and new good codes are obtained from the structural results. Further, in this regards few open problems are listed below:
\begin{enumerate}
    \item  The self-dual $\mathbb{Z}_{p}\mathbb{Z}_{p}[u,v]$-additive cyclic codes would be interesting in view of \cite{Aydogdu17self,Aydogdu19self}.
    \item The quantum codes from $\mathbb{Z}_{p}\mathbb{Z}_{p}[u,v]$-additive cyclic and constacyclic codes under the CSS construction might produce significantly new quantum codes similar to \cite{Diao18,Diao19}.
\end{enumerate}

\section*{Acknowledgement}
The authors are thankful to the Department of Science and Technology (DST) (under project MTR/2022/001052, vide Diary No. SERB/F/8787/ 2022-2023 dated 29 December, 2022) for financial supports and Indian Institute of Technology Patna for providing the research facilities. We would like to thank Prof. Patrick Sol\`{e} (University Aix-Marseille, Marseille, France) for his careful reading and suggestions.
\section*{Declarations}
\textbf{Data Availability Statement}: The authors declare that [the/all other] data supporting the findings of this study are available within the article. \\
\textbf{Competing interests}: The authors declare that there is no conflict of interest regarding the publication of this manuscript.\\

\end{document}